\documentclass[superscriptaddress,aps,prd,twocolumn,showpacs,nofootinbib]{revtex4-1}
\usepackage{etex}
\usepackage{amsmath,amssymb,amsthm}
\usepackage[utf8x]{inputenc}
\usepackage[colorlinks=true,citecolor=blue,urlcolor=blue]{hyperref}
\usepackage[pdftex]{graphicx}  
\usepackage{verbatim}   
\usepackage{xcolor}      
\usepackage{subfigure}  
\usepackage{hyperref}   
\usepackage{mathrsfs}
\usepackage{booktabs}
\usepackage[font=footnotesize,labelfont=bf]{caption} 
\usepackage{wrapfig}
\begin{document}
\title{Evolution of axial perturbations in space-time of a non-rotating uncharged primordial black hole }
\author{Arnab Sarkar} \email{arnabsarkar@bose.res.in, arnab.sarkar14@gmail.com}
\affiliation{Department of Astrophysics and High energy Physics, 
S. N. Bose National Centre for Basic Sciences, JD Block, Sector III, Salt lake city, Kolkata-700106, India.}
\author{Sabiruddin Molla}\email{sabiruddinmolla111@gmail.com}
\affiliation{St. Xavier's College (Autonomous), 30 Park St., Mullick Bazar, Park Street area, Kolkata-700016, India. }
\author{K. Rajesh Nayak }\email{rajesh@iiserkol.ac.in}
\affiliation{Indian Institute of Science Education and Research, Mohanpur, West Bengal-741246, India.}
\date{\today}
\begin{abstract}
\begin{center}
\textbf{Abstract}
\end{center}
\begin{small}
We derive the equation governing the axial-perturbations in the space-time of a non-rotating uncharged primordial black hole (PBH), produced in early Universe, whose metric is taken as the generalized McVittie metric. The generalized McVittie metric is a cosmological black hole metric, proposed by V. Faraoni and A. Jacques in 2007 \cite{Faraoni}. This describes the space-time of a Schwarzschild black hole embedded in FLRW-Universe, while allowing its mass-change.
Our derivation has basic similarities with the procedure of derivation of S. Chandrasekhar, for deriving the Regge-Wheeler equation for Schwarzschild metric \cite{Chandrasekhar} ; but it has some distinct differences with that due to the complexity and time-dependency of the generalized McVittie metric.
We show that after applying some approximations which are very well valid in the early radiation-dominated Universe, the overall equation governing the axial perturbations can be separated into radial and angular parts, among which the radial part is the intended one, as the angular part is identical to the case of Schwarzschild metric as expected.
We identify the potential from the Schr\"{o}dinger-like format of the equation and draw some physical interpretation from it. 
\end{small}
\end{abstract}
\maketitle
\vspace{0.5cm}
\begin{small}
\section{Introduction }\label{section-1}
Perturbations in black hole space-times has been an interesting topic of research for the last few decades. At present, after the first direct detection of gravitational waves and subsequent series of detections of gravitational waves from various similar sources \cite{LIGO-1, LIGO-VIRGO-2, LIGO-VIRGO-3, LIGO-VIRGO-4, LIGO-VIRGO-5, LIGO-VIRGO-6}, it is even more important for observational estimation of different parameters of gravitational wave sources, related to black holes, specially newly born black holes in `ring-down phase' after merging of a binary of black holes. Black hole perturbation theory is one of the most important tools for accurately determining characteristics of this type of gravitational wave sources, through gravitational wave astronomy. The perturbations in a black hole's space-time can be created by various means e.g. (i) perturbations can be generated in the space-time of a black hole due to inspiralling motion of a comparatively very smaller particle (i.e. test-particle) around it ; (ii) the merging of two black holes in a binary creates perturbations in the newly born resultant black hole ; (iii) infall or interaction of gravitational waves from other sources can also create perturbations in a black hole's space-time etc.. In each of these different cases, black hole perturbation theory is pivotal to study the evolution of perturbations in the space-time of the black hole.  \\
In 1957, Tullio Regge and John A. Wheeler derived the equation describing the behaviour or evolution of axial-perturbations in Schwarzschild metric \cite{Regge-Wheeler} and using this, they studied the stability of the Schwarzschild geometry, when it is subjected to small non-spherical perturbations. This equation is known as `Regge-Wheeler' equation after their name. This can be called effectively the birth of `black hole perturbation theory'.
 Later S. Chandrasekhar derived the same equation in a different procedure and in a more general way\cite{Chandrasekhar}. \\
 Later in 1970, Zerilli extended the analysis to polar-perturbations in the Schwarzschild space-time \cite{Zerilli-1, Zerilli-2}. He showed that the equations governing the perturbations can be expressed as a pair of Schr\"{o}dinger-like equations and he applied the formalism to study the gravitational radiation emitted by infalling test-particles into black holes.    \\ 
The Regge-Wheeler and the Zerilli equations or perturbation techniques developed by them, and similar counterparts for other different types of black holes, paved the way to investigate the stability of  perturbations for certain modes of vibration of the black holes. Instability or stability of perturbations mean whether the perturbations grow with time, thereby becoming too large to be handled by the linear perturbation theory or those decay gradually with time respectively. C. V. Vishveshwara analyzed the stability of the Schwarzschild black hole, through a numerical experiment, by slightly perturbing it with an infalling wave packet, thereby observing the scattered wave \cite{Vishveshwara}. He found that the scattered wave is a sum of damped sinusoids, whose frequencies and damping times are the \textit{`quasinormal modes'} i.e. the charateristic modes of free vibration of the black hole. The damping implies that the concerned black hole is stable i.e. returns into a stationary state after being perturbed. The outcome can be different for different types of black hole metrics and even can be different for same black hole metric with different environments, i.e. presence of some other matter near it.  
\par In the present work, we derive the equation governing axial-perturbations in the space-time of a cosmological black hole, called `generalized McVittie metric', proposed by V. Faraoni and A. Jacques in 2007 \cite{Faraoni}. This is a generalization over the original `McVittie metric', given by G. C. McVittie \cite{McVittie}. This generalized McVittie metric describes the space-time geometry of a Schwarzschild black hole embedded in FLRW-Universe, while allows change of mass of the black hole.\\
 The reason for the choice of this metric in our work, for describing the space-time around non-rotating uncharged primordial black holes (PBHs) has been described in the section \ref{section-2}. PBHs are thought to be produced in early Universe by direct gravitational collapse of regions with sufficient over-density after the horizon re-entry. Our main motivation is to study the evolution of perturbations in the space-time of a PBH in early radiation-dominated Universe, which has two distinct differences in comparison with the space-time of any astrophysical black hole in the  present-era or late-time Universe. These differences are : (i) most of the PBHs in early Universe were subjected to rapid rate of change of mass due to either spherical accretion of the surrounding high-density radiation \cite{Majumdar_et_al, Upadhyay_et_al, Custodio_et_al, Custodio_et_al_2} or due to Hawking evaporation and (ii) the effect of expansion of the Universe on the space-time around a PBH was significant due to the robust value of Hubble-parameter at that early era, in comparison to the late Universe. Due to these two differences, it is expected that the evolution of metric-perturbations around a PBH in early radiation-dominated Universe would be different than that of around an astrophysical black hole in the Universe of present era and that can not be explained by the usual equations viz. `Regge-Wheeler equations' and `Zerilli equations' for Schwarzschild metric or `Tuekolsky master equations' for Kerr metric. \\    
Propagation of gravitational waves in an expanding background in presence of a point-mass i.e. in Newtonian McVittie background has been investigated recently \cite{Antoniou_et_al} and it has been shown that the point-mass increases the amplitude of the gravitational wave, while decreases its frequency, relative to an observer placed at infinity. However, a point-mass can hardly describe the more realistic scenario of the space-time around a  PBH. Hence, exploration of the more generalized and relevant case of the generalized McVittie metric is necessary.   
\\
In deriving the desired equation, which governs the axial perturbations in the generalized McVittie metric, we have followed the basic procedure similar to that used by S. Chadrasekhar for deriving the Regge-Wheeler equation for the Schwarzschild metric. Yet, there are some fundamental differences. One of the main differences is that the unperturbed parameters, appearing in metric-coefficients of the generalized McVittie metric, are time-dependent. This is why the usual way of Fourier-transforming only the perturbations from time-space to frequency-space may not work here and this would ultimately give the intended equation as spatio-temporal equation, which would be although mathematically  correct, but physically very hard to solve and interpret. For this reason, we employ Fourier-transformation of the overall perturbation equations. Then, applying various results of the `Convolution theorem' for Fourier-transform of product of two functions and using some approximations, we get the equation in desired form.
After transforming the equation in Schr\"{o}dinger-like form, we identify the potential from this form of the equation and then analyze some important aspects of it.    
\par
The paper is organized as follows : in the section \ref{section-2}, we describe the reason for our choice of the generalized McVittie metric for describing the space-time around a non-rotating, uncharged PBH, which is subjected to change of mass in the early radiation-dominated Universe. In this section we also describe the convenient forms of the line-element of this metric in different coordinates. In the section \ref{section-3}, we describe the derivation of the equation governing the axial-perturbations in the generalized McVittie metric to a certain level, and in the section \ref{section-4} we simplify that equation by applying some approximations. We also separate the radial and angular parts of the equation with the application of separation of variables technique, where the radial part is the intended equation, as the angular part is identical to that of the Schwarzschild metric. Furthermore, in this section we transform the equation in a form free of first-order derivative term i.e. the Schr\"{o}dinger-like form, from which we identify the potential to draw some physical interpretation from it.
 In the last section \ref{section-6}, we give final remarks of our work. Besides these, there are five Appendices to complement and clarify various aspects of this work. \\
It is to be noted that we have mainly followed the natural system of units in the analytical calculations, where `G' and `c' are set to 1 or omitted. But, during discussing some numerical-orders, we use G and c in their necessary places in the expressions.    
\section{The reason for choice of the generalized McVittie metric and its forms in different coordinate systems  }
\label{section-2}
To derive the equation governing the axial-perturbations in space-time around a PBH in early Universe, first of all the correct metric describing the space-time is to be choosen. As in the early Universe, when the PBHs were produced and the era in which we are interested, the rate of expansion of the Universe i.e. the Hubble-parameter was very high, in comparison with the present era. Therefore, this effect of robust cosmological expansion on the local space-time around the PBHs can not be neglected. There were series of efforts to describe this effect and hence to get a resultant metric, which can describe the space-time of a black hole embedded in an expanding Universe, 
when the effect of expansion of the Universe on the space-time around the black hole is significant. This type of metrics are usually referred as `cosmological black hole metrics'.\\
One of the first attempts was by McVittie \cite{McVittie}. The solution, given by him, is known as `McVittie metric'. But, this metric can describe the intended space-time, provided the mass of the black hole is not changing with time. However, in the case of PBHs in early Universe, this condition of constancy of mass would be impractical because, there was highly dense radiation almost everywhere in the radiation-dominated era and hence this high-density radiation was subject to spherical accretion by the PBHs, leading to the growth of masses of the PBHs. Also, as PBHs spanned an enormous mass-range, from the end of inflation ($10^{-32} \, s$) up to the big bang nucleosynthesis ($\sim 1 \, s$) \cite{Carr-3} and a large fraction of the PBHs, created in early Universe, were of smaller mass than the Solar-mass ; Hawking-evaporation was a prominent phenomenon for those.\\ 
Some PBHs were in the mass-range such that for them the rate of loss of mass due to Hawking radiation should be very dominant, such that the rate of gain of mass due to spherical accretion of the surrounding radiation would be insignificant in comparison to the mass-loss rate due to Hawking evaporation. Again, there were some PBHs in the mass range such that for them the rate of loss of mass due to Hawking evaporation was negligible with respect to the rate of mass-gain by spherical accretion of high-density radiation. 
 So, there is no doubt that most of the PBHs were in the mass range such that they were undergone mass change with time, whether was it gain of mass or loss of mass. So, the condition of constancy of mass is not justified at all for those PBHs. \\
 That is why there were more works following the work of McVittie, trying to give a more generalized solution where there is not any restriction on the change of mass. One of the metrics, which has been subject of much interest, is the `Schwarschild-De Sitter metric'. When the background metric is chosen to be De Sitter, the McVittie metric reduces to Schwarzschild-De Sitter metric. The `Schwarschild-Anti De Sitter metric' is also of similar interest. But, these are vacuum solutions of Einstein's equations and hence these do not allow the mass-change of PBHs due to accretion of surrounding radiation, in the early rdaiation-dominated era of the Universe. \\     
 Another two attempts, to describe such black hole metrics, are the Sultana-Dyer solution \cite{Sultana-Dyer} and MaClure-Dyer solution \cite{McClure-Dyer}. 
  But, each of these has their own shortcomings to describe the metric of a black hole embedded in an expanding FLRW-Universe. Their deficiencies have been briefly described  in references \cite{Faraoni, Faraoni2}. 
   \\
 In 2007, V. Faraoni and A. Jacques \cite{Faraoni} gave a new solution, better to say they proposed a more generalized extension of the McVittie metric. They showed that it can describe mass-change of a Schwarzschild black hole embedded in an expanding FLRW-Universe, due to spherically accreting surrounding cosmic-fluid (viz. radiation in this case). 
 They called it the `Generalized McVittie metric'. More explanation and emphasis on this metric was given in a subsequent work \cite{GaoChen}.\\
 For the present purpose, we choose this new metric, as it does not require any imposed condition on the change of mass of the black holes, while also it does not seem to have any major theoretical drawbacks.\\
  This metric is written in isotropic coordinates as:
 \begin{equation} \label{1.1}
 ds^{2} = -\frac{\mathcal{B}^{2}(t,r)}{\mathcal{A}^{2}(t,r)} dt^{2} + a^{2}(t) \mathcal{A}^{4}(t,r)(dr^{2}+ r^{2}d\Omega^{2}) \, , 
\end{equation}  
where, $r $ and $t $ are respectively the isotropic radial and time coordinates. The quantities $ \mathcal{A} $ and $\mathcal{B} $ are given by :
\begin{center}
 ${\displaystyle \mathcal{A}(t,r) = 1+ \frac{m(t)}{2r}}$ and ${\displaystyle \mathcal{B}(t,r) = 1-\frac{m(t)}{2r}  }$, 
\end{center}
where $m(t) $ is the mass of the PBH and it is time-varying. 
$d\Omega^{2} = r^{2}(d\theta^{2} + Sin^{2} \theta d\phi^{2}) $ is the usual angular part. \\ 
For convenience, the authors of reference \cite{Faraoni} introduced the quantity :
\begin{equation} \label{1.2}
\mathcal{C} = \Big( \frac{\dot{a}}{a} + \frac{\dot{m}}{r \mathcal{A}}   \Big) = \frac{\dot{M}_{H}}{M_{H}}- \frac{\dot{m}}{m}\frac{\mathcal{B}}{\mathcal{A}} \, , 
\end{equation}
where $M_{H}$ represents the `Hawking-Hayward quasi-local mass' of the black hole and $a(t)$ is the scale-factor of the background FLRW-Universe. The `dot' denotes differentiation with respect to the isotropic time coordinate t. \\ 
This metric \ref{1.1} can be transformed to a Schwarzschild-like coordinate system. C. Gao et al have described this process of transforming the metric into a Schwarzschild-like form, in their work \cite{GaoChen}. Yet we are mentioning it briefly as the definitions and inter-relations of corresponding coordinates are required in our work.   
First by defining the areal radius
\begin{equation} \label{1.3}
   \tilde{r} = r \Big( 1+ \frac{M_{H}(t)}{2ra(t)} \Big)^{2} 
\end{equation}
 and using $ m(t) = M_{H} (t)/a(t)$,
and then introducing the co-moving radial coordinate $R= a\tilde{r}$, in terms of which ${\displaystyle  d\tilde{r} = \frac{dR}{a}- H \tilde{r} dt }$, equation \ref{1.1} is turned into the Painleve-Gullstrand form :
\begin{equation}  \label{1.5}
\begin{aligned}
 ds^{2} = - \left\lbrace 1-\frac{2M_{H}}{R} -\frac{\Big( HR+ \dot{m} a \sqrt{\frac{\tilde{r}}{r}}    \Big)^{2}}{1-\frac{2M_{H}}{R}}  \right\rbrace dt^{2} +
 \\
\frac{1}{1-\frac{2M_{H}}{R}} dR^{2} +R^{2} d\Omega^{2} - \frac{2}{1-\frac{2M_{H}}{R}} \left\lbrace HR + \dot{m} a \sqrt{\frac{\tilde{r}}{r}} \right\rbrace dt dR \, . 
\end{aligned}
\end{equation}
Further setting 
\begin{equation} \label{1.6}
\displaystyle  A(t,r) = {1-\frac{2M_{H}}{R}}  \, , 
\end{equation}
 \begin{equation} \label{1.7}
  C(t,r) = HR + \dot{m} a \sqrt{\frac{\tilde{r}}{r}} \, , 
\end{equation}  
   and defining the time coordinate $\overline{t}$ as 
\begin{equation} \label{1.8}
d\overline{t} = \frac{1}{F} \Big( dt + \frac{C}{A^{2}-C^{2}} dR  \Big)   \,  , 
\end{equation}  
where $F(t,r)$ is an integrating-factor that makes $d\overline{t}$ an exact-differential ,
one gets 
\begin{equation} \label{1.9} 
\begin{aligned}
ds^{2} = -\frac{(A^{2}- C^{2})}{A} \left\lbrace  F^{2}d\overline{t}^{2} + \frac{C^{2}dR^{2}}{(A^{2}- C^{2})^{2}} - \frac{2FB}{(A^{2}- C^{2})} d\overline{t} dR \right\rbrace 
\\
+\frac{dR^{2}}{A} + R^{2} d\Omega^{2} - \frac{2C}{A} dR \Big( Fd\overline{t} -\frac{C}{(A^{2}- C^{2})} dR \Big) \, . 
\end{aligned}
\end{equation}
The cross terms containing $dRd\overline{t}$ cancel out and the squared line-element in the new `Nolan-gauge' becomes 
\begin{equation} \label{1.10}
\begin{aligned}
{\displaystyle  
ds^{2} = -A \Big( 1 - \frac{C^{2}}{A^{2}} \Big) F^{2} d\overline{t}^{2} 
} 
+ A^{-1} \Big(  1 - \frac{C^{2}}{A^{2}} \Big) ^{-1} dR^{2}+ R^{2} d\Omega^{2}  \,  .
\end{aligned}
\end{equation}    
We are using this form \ref{1.10} in our work. There may be a question that why we choose to work with this form \ref{1.10} in `Nolan gauge', which is a Schwarzschild-like form, instead of using the form \ref{1.1} in isotropic coordinate system, despite the fact that working in isotropic-coordinate system seems to be comparatively simpler. The main reason is that we want to utilize the symmetry of the form \ref{1.10} viz. if we see the time coordinate as $\tilde{t} $ such that $d\tilde{t} = F d\overline{t} $, then $g_{00} = - g_{11}^{-1}$ , where the indices `0' and `1' stand for temporal and radial coordinate respectively. Furthermore, the isotropic radial coordinate $ r $ does not always faithfully represent radial distances. This fact is very disturbing while interpreting the results with practical cases.   
\section{ Derivation of equations describing the axial perturbations in Generalized McVittie metric }\label{section-3}
The square of line-element of a generalized metric can be written as : 
\begin{equation} \label{4.1}
\begin{aligned}
ds^{2} =- e^{2 \nu} d\tau^{2} + e^{2 \psi } (d\phi - \omega d\tau - q_{2} dx_{2} - q_{3} dx_{3})^{2} + e^{2 \mu_{2}}dx_{2} ^{2}
\\
 +  e^{2 \mu_{3}}dx_{3} ^{2} \, , 
\end{aligned}
\end{equation}
where the quantities $ \nu, \psi, \mu_{2}, \mu_{3}, \omega, q_{2}, q_{3} $ are functions of the coordinates $ \tau, x_{2}, x_{3}$. Now we compare the form of the metric given in the equation \ref{4.1} with the generalized McVittie metric proposed by V. Faraoni et al \cite{Faraoni} in Nolan Gauge, as given in equation \ref{1.10}, with the corresponding coordinates being $ \tau \equiv \overline{t} \, , x_{2} \equiv R \, , x_{3} \equiv \theta $ and $ \phi\equiv \phi  $ (We follow the index designation : $0, 1, 2, 3 $ stand for respectively $\overline{t}, \phi, R, \theta $). Comparing the metric in equation \ref{4.1} with the 
 generalized McVittie metric in Nolan Gauge, as given in equation \ref{1.10}, we see that the coefficients $ \nu, \psi, \mu_{2}, \mu_{3}, \omega, q_{2}, q_{3}  $, in case of generalized McVittie metric are given by the set of equations :
\begin{eqnarray} \label{4.2a}
 -e^{2\nu} + e^{2 \psi} \omega^{2} = -\Big( 1- \frac{2 M_{H}}{R}  \Big) \Big(  1- \frac{C^{2}}{A^{2}}  \Big)F^{2} \, , \\
 \label{4.2b}
 e^{2\psi} q_{2}^{2} + e^{2 \mu_{2}}  = \Big( 1- \frac{2 M_{H}}{R}  \Big)^{-1} \Big(  1- \frac{C^{2}}{A^{2}}  \Big)^{-1} \, , \\
 \label{4.2c}
 (e^{2\psi} q_{3}^{2} + e^{2 \mu_{3}} ) = R^{2} \, , \\
 \label{4.2d}
 e^{2\psi} = R^{2} Sin^{2} \theta \, .
\end{eqnarray}
\footnote{One issue is to be noted here that the metric signatures of the metric in form \ref{4.1} is opposite to that used in reference \cite{Chandrasekhar}. For this reason in each of the expressions we are using here, there will be a negative multiplicity with $ e^{2\alpha}$ ($ \alpha = \nu, \psi, \mu_{2} , \mu_{3} $). While except this sign change, there would not be any other change in the expressions of the Ricci tensors.  }
While the absence of cross-terms (i.e. terms with $dx_{i}dx_{j}, \, d\tau dx_{i}, dx_{i}d\phi$ etc.) in the metric \ref{1.10} indicate that the zeroth order or unperturbed values of the coefficients causing the cross terms are zero \textit{viz.} comparing with the metric given in equation \ref{4.1} : $ \omega = 0 \, , q_{2} = 0 \, , q_{3} = 0 $. \\ Then solving the above set of equations \ref{4.2a} to \ref{4.2d} for the coefficients $ \nu, \psi, \mu_{2} $ and $ \mu_{3} $, we obtain :
\begin{eqnarray} \label{4.3}
\nu = \frac{1}{2} ln \left\lbrace  \Big( 1- \frac{2 M_{H}}{R} \Big) \Big(  1- \frac{C^{2}}{A^{2}}  \Big)F^{2} \right\rbrace  \, ,  \\
\mu_{2} = \frac{1}{2} ln\left\lbrace \Big( 1- \frac{2 M_{H}}{R} \Big)^{-1} \Big(  1- \frac{C^{2}}{A^{2}}  \Big)^{-1}    \right\rbrace  \, ,   \\
\mu_{3} = \frac{1}{2} ln(R^{2}) \, ,  \\
\psi = \frac{1}{2} ln(R^{2} Sin^{2} \theta ) \, . 
\end{eqnarray}
Now, we have to get the expressions of the Ricci tensor components $ R_{12}$ and $R_{13} $, upto the first order perturbations of the quantities $ \omega $, $q_{2} $ and $q_{3} $. Here we denote the first order or linear perturbations in $\omega \,  ,q_{2} $ and $q_{3}  $ as $\delta\omega \, ,\delta q_{2}$  and $\delta q_{3}$ respectively. But, as we have already stated that in this case the background or zeroth order values of these quantities are zero : $\omega= q_{2}= q_{3}= 0  $, hence the overall quantities can be given by $\delta\omega \, ,\delta q_{2}$  and $\delta q_{3}$. The axial perturbations (as called in reference \cite{Chandrasekhar}) are characterized by the non-zero values of $\delta\omega \, ,\delta q_{2}$  and $\delta q_{3}$. Any general perturbation of this metric \ref{4.1} would generate the perturbations $\delta\omega \, ,\delta q_{2}$  and $\delta q_{3}$, with zero   unperturbed values for the case without any cross-terms. While the non-zero unperturbed values of the quantities $ \nu\, , \psi\, , \mu_{2}\, , \mu_{3}$ would experience first order perturbations $ \delta\nu \, , \delta\psi \, , \delta\mu_{2} \,$ and $\delta\mu_{3}$ respectively. But, as in the case of Schwarzschild metric, in this case of generalized McVittie metric too, these two sets of perturbations have completely different effects. 
  As argued in reference \cite{Chandrasekhar}, the set of perturbations  $\delta\omega \, ,\delta q_{2}$  and $\delta q_{3}$ induce a dragging of the inertial frame thereby imparting a rotation on the black hole. But, the other set has no such rotational effects. For this reason they are respectively called as Axial and Polar perturbations in reference \cite{Chandrasekhar}, on the basis of effect of sign-reversal of $ \phi $ on the metric. On the basis of this different behaviour, we can physically interpret that they must decouple .\\
For this reason instead of getting the equations, 
which govern all the perturbations in a metric, in detail i.e. where both the sets of perturbations will be present, we can extract the part containing the set $\delta\omega \, ,\delta q_{2}$  and $\delta q_{3}$ first and then the part containing the other. The part containing the set of perturbations $\delta\omega \, ,\delta q_{2}$  and $\delta q_{3}$, will contain the background values of quantities  $ \nu\, , \psi\, , \mu_{2}\, , \mu_{3}$.   \\
 Following reference \cite{Chandrasekhar}, we use the definitions : 
\begin{eqnarray} \label{4.4}
Q_{AB} = \delta q_{A, B }- \delta q_{B, A }  \,  ,   \\
Q_{A0} = \delta q_{A, 0 }- \delta \omega_{,A }  \, ,
\end{eqnarray}
where the indices A, B = 2,3 in this case. \\
\footnote{\textit{
Our notation of representing the Ricci tensor components, upto first order perturbation of the quantities $ \omega \, , q_{2} $ and $q_{3} $ is : $R_{ij} + \delta_{\omega,q_{2},q_{3}}  R_{ij}  $.
}
}
Using the expressions given in the reference \cite{Chandrasekhar}, the components of Ricci tensor $ R_{12} + \delta_{\omega,q_{2},q_{3}}  R_{12}$ and $ R_{13}+\delta_{\omega,q_{2},q_{3}}  R_{13} $ (these give the first order perturbations only, because the background values of these Ricci tensor components $R_{12} $ and $ R_{13} $ are zero ;) in our case are given by :
\begin{equation} \label{4.5}
\begin{aligned}
R_{12} + \delta_{\omega,q_{2},q_{3}}  R_{12} = \frac{1}{2}e^{-2\psi} (e^{-2 \nu} e^{-2 \mu_{3}})^{1/2}
\\
 [ ((e^{3\psi + \nu -(\mu_{2} + \mu_{3})}) Q_{32})_{,3} - ((e^{3\psi - \nu -\mu_{2} + \mu_{3}}) Q_{02})_{,0} ]  \, , 
\end{aligned}  
\end{equation}
\begin{center}
and
\end{center}
\begin{equation} \label{4.6}
\begin{aligned}
R_{13} + \delta_{\omega,q_{2},q_{3}}  R_{13}  = \frac{1}{2}e^{-2\psi} (e^{-2 \nu} e^{-2 \mu_{2}})^{1/2}
\\
 [ ((e^{3\psi + \nu -(\mu_{2} + \mu_{3})}) Q_{23})_{,2} - ((e^{3\psi - \nu +\mu_{2} - \mu_{3})}) Q_{03})_{,0} ] \,  . 
\end{aligned}
\end{equation}
Before proceeding we need the values of the perturbation to the metric components $ g_{12}$ and $ g_{13}$ i.e. $\delta_{\omega,q_{2},q_{3}}g_{12} $ and $\delta_{\omega,q_{2},q_{3}}g_{13}  $. It is to be noted 
that the metric is given in covariant form in the line-element \ref{4.1}. 
So, the perturbation to the metric components w.r.t. $ \omega \, ,q_{2} \, $ and $q_{3} $ are given by :
$ \delta_{\omega,q_{2},q_{3}} g_{12} = -e^{2 \psi}  \delta q_{2} $ and $ \delta_{\omega,q_{2},q_{3}} g_{13} = -e^{2 \psi}  \delta q_{3} $ . 
 \\
Now, we start getting the equations describing the axial perturbations for the generalized McVittie metric. The origin of the equations has been discussed in APPENDIX 2 \ref{BasicPertEqn}. \\
The equation $ \delta_{\omega,q_{2},q_{3}}R_{12} - \frac{1}{2}  (\delta_{\omega,q_{2},q_{3}} g_{12} )  \mathcal{R} = 0 $ is :
\begin{equation} \label{4.7}
\begin{aligned}
-\frac{1}{2}e^{-2 \psi}(e^{-2\nu} e^{-2 \mu_{3}})^{1/2} [ ((e^{3\psi + \nu -(\mu_{2} + \mu_{3})}) Q_{23})_{,3} +
\\
 ((e^{3\psi - \nu -\mu_{2} + \mu_{3}}) Q_{02})_{,0}   ] = 
\frac{1}{2} (- e^{2\psi} \delta q_{2}) \mathcal{R}   \,  ,
\end{aligned}
\end{equation}
and the equation $ \delta_{\omega,q_{2},q_{3}}R_{13} - \frac{1}{2}(\delta_{\omega,q_{2},q_{3}} g_{13}) \mathcal{R} = 0 $ is :
\begin{equation} \label{4.8}
\begin{aligned}
\frac{1}{2}e^{-2 \psi}(e^{-2\nu} e^{-2 \mu_{2}})^{1/2} ((e^{3\psi + \nu -(\mu_{2} + \mu_{3})}) Q_{23})_{,2} -
\\
   ((e^{3\psi - \nu +\mu_{2} - \mu_{3})}) Q_{03})_{,0} = \frac{1}{2} (- e^{2\psi} \delta q_{3}) \mathcal{R}  
\, . 
\end{aligned}
\end{equation}
The $\mathcal{R}$ in the above equations is the Ricci-scalar. Although we shall see later, that $\mathcal{R} $ vanishes in the scenario of our interest, yet we keep the $\mathcal{R} $ in the equations to a certain stage, so that the appearance of the $\mathcal{R} $ in the desired final equation can be checked. In any more general case, where the $\mathcal{R}$ does not vanish, this may be utilized.    \\
After shifting the factor $e^{-2\psi} $ from LHS to RHS of the equations \ref{4.7} and \ref{4.8}, we write these as respectively :
\begin{equation} \label{4.7a}
\begin{aligned}
-\frac{1}{2} (e^{-2\nu} e^{-2 \mu_{3}})^{1/2} [ ((e^{3\psi + \nu -(\mu_{2} + \mu_{3})}) Q_{23})_{,3} +
\\
 ((e^{3\psi - \nu -\mu_{2} + \mu_{3}}) Q_{02})_{,0}   ] = 
\frac{1}{2} e^{4 \psi} (-  \delta q_{2}) \mathcal{R} \, , 
\end{aligned}
\end{equation}
and 
\begin{equation} \label{4.8a}
\begin{aligned}
\frac{1}{2} (e^{-2\nu} e^{-2 \mu_{2}})^{1/2} ((e^{3\psi + \nu -(\mu_{2} + \mu_{3})}) Q_{23})_{,2} -
\\
   ((e^{3\psi - \nu +\mu_{2} - \mu_{3})}) Q_{03})_{,0} = \frac{1}{2} e^{4 \psi} (-  \delta q_{3}) \mathcal{R} \, . 
\end{aligned}
\end{equation}
Before proceeding we define some quantities for brevity and compactness of the upcoming equations, as was defined in reference \cite{Chandrasekhar} for Schwarzschild metric, as below :
\begin{equation} \label{4.9}
Q(\overline{t}, R, \theta )= Q_{23}\Delta Sin^{3} \theta  \, , 
\end{equation}
where the quantity $ \Delta $ is given by :
\begin{equation} \label{4.10}
\Delta = R^{2} - 2 M_{H} R \, . 
\end{equation}
Hence,
\begin{equation}   \label{4.11}
\Big( 1- \frac{2 M_{H}}{R}    \Big)   =  \Big(\frac{R^{2}- 2 M_{H} R}{R^{2}}  \Big) \equiv \frac{\Delta}{R^{2}}   \, . 
\end{equation}
Following expressions of the unperturbed coefficients present in the metric of form \ref{4.1} for the metric \ref{1.10} are useful :
\begin{center}
${ \displaystyle  e^{3\psi} = R^{3} \, Sin^{3} \theta    \,   ,     }$  \\
${ \displaystyle  e^{\nu} =  \Big( 1 - \frac{2 M_{H}}{R}   \Big)^{1/2} \Big( 1 - \frac{C^{2}}{A^{2}}  \Big)^{1/2} F            \, , }$                     \\
${ \displaystyle  e^{-\mu_{3}} = R^{-1}    \, ,      }$             \\
${ \displaystyle  e^{\mu_{2}} = \Big(  1 - \frac{2 M_{H}}{R}  \Big)^{-1/2}  \Big(   1 - \frac{C^{2}}{A^{2}} \Big)^{-1/2} \, . }$   
\end{center}
Now, substituting the expressions of $ e^{2\alpha}$ ($ \alpha = \nu, \psi, \mu_{2} , \mu_{3} $), in the  equations \ref{4.7a} and \ref{4.8a}, we obtain the quantities present in these equations for the metric given in equation \ref{1.10}. 
According to the convention and notation of S. Chandrasekhar in reference \cite{Chandrasekhar}, the quantity $ Q_{02}$ is given by :
\begin{equation} \label{4.14}
Q_{02}= -Q_{20} = (\delta\omega_{,2} - \delta q_{2,0})  \, . 
\end{equation}
From now, we shall denote the quantity $F  \Big(   1 - \frac{C^{2}}{A^{2}} \Big)  $ as $\lambda $ for brevity. 
Hence, the equation \ref{4.7a} in our case  
(rearranging it in a form of our convenience) can be written as :
\begin{equation} \label{4.15}
\begin{aligned}
\frac{F \lambda }{R^{4}\, Sin^{3}\theta}  \frac{\partial Q }{\partial \theta}  = 
F^{2} R \, Sin \, \theta \, \Big( 1- \frac{2 M_{H}}{R} \Big)^{\frac{1}{2}} \Big(  \frac{\lambda}{F} \Big)^{\frac{1}{2}} \mathcal{R} \delta q_{2} 
\\
 -\Big( \frac{\partial^{2} \delta \omega}{\partial \overline{t} \partial R}- \frac{\partial^{2} \delta q_{2}}{\partial \overline{t}^{2}} \Big) -  \Big( \frac{\partial \delta \omega}{\partial R} - \frac{\partial \delta q_{2}}{\partial \overline{t}} \Big)                                 
   \frac{F}{R^{4}}\frac{\partial}{\partial \overline{t}} \Big( \frac{R^{4}}{F}  \Big) \, . 
 \end{aligned}  
\end{equation} 
On the other hand, equation \ref{4.8a} in our case, for the metric given in equation \ref{1.10}, can be conveniently written as below :
\begin{equation} \label{4.19}
\begin{aligned}
\frac{\Delta}{R^{4}\, Sin^{3}\theta} \left(   \frac{\partial Q}{\partial R} + \frac{ Q}{\lambda}   \frac{\partial \lambda}{\partial R}   \right)
 - \frac{1}{\lambda^{2}}  \frac{\partial}{\partial \overline{t}} \Big( \frac{\partial \delta\omega}{\partial \theta } - \frac{\partial \delta q_{3}}{\partial \overline{ t}}  \Big) 
 \\
 - \frac{\Delta}{R^{4} \lambda \, Sin^{3}\theta} 
 \Big( \frac{\partial \delta\omega}{\partial \theta } - \frac{\partial \delta q_{3}}{\partial \overline{ t}}  \Big)  \frac{\partial}{\partial \overline{t}} \Big(   \frac{R^{4}\, Sin^{3}\theta}{\Delta \, F} \Big(\frac{\lambda}{F}\Big)^{-1} \Big)  
 \\
 = - \delta q_{3} \Delta \, Sin \, \theta  \, \Big(\frac{\lambda}{F}\Big)^{-1} \mathcal{R} \, . 
 \end{aligned}
\end{equation}
At this point, we shall face difficulty 
if we assume that the perturbations have time-dependence proportional to $e^{i \sigma \overline{t}} $, where $\sigma $ is a constant, or, in other way to say, if we take the Fourier-transform of the perturbations from $\overline{t}$-space to $ \sigma $-space. Then the resulting terms will not serve the purpose. This is because in the coordinate system in Nolan-gauge, which we are using, the time-coordinate $\overline{t} $ is a function of both the isotropic radial and time coordinates viz. $ r $ and $ t $. Also, the radial coordinate $R$ and the temporal coordinate $\overline{t} $ are inter-related. For the same reason, in the term $\frac{\partial^{2} \delta \omega}{\partial \overline{t} \partial R }$, the partial derivatives w.r.t. $R $ and $\overline{t} $ can not be commuted.\\
So, to avoid this complication in the calculations, we shall convert the partial derivatives w.r.t. $R $ and $\overline{t} $ into those w.r.t. $ r $ and $ t $ respectively. \\
Expressing the following double-partial derivatives of the perturbations in terms of partial derivatives w.r.t. $r $ and $t $ (using the formula given in APPENDIX-3 i.e. section  \ref{RelationsforGMcM}), we obtain :
\begin{equation} \label{A1}
\begin{aligned} 
\frac{\partial^{2} \delta \omega }{\partial \overline{t } \partial R} 
=  \left\lbrace \lambda   \frac{\partial }{\partial t}  \Big(  a^{-1}\Big( 1 - \frac{M_{H}^{2}}{ 4 a^{2} r^{2} } \Big)^{-1} \Big) \right\rbrace \frac{\partial \delta \omega}{\partial r } +
\\
 \left\lbrace  \lambda \Big(  a^{-1}\Big( 1 - \frac{M_{H}^{2}}{ 4 a^{2} r^{2} } \Big)^{-1} \Big) \right\rbrace  \frac{\partial^{2} \delta \omega}{\partial t \partial r }   \,  , 
\end{aligned} 
\end{equation}
\begin{center}
and 
\end{center}
\begin{equation}  \label{A2}
\begin{aligned}  
\frac{ \partial^{2} \delta q_{2}}{ \partial \overline{t}^{2}}  
=  \left( \lambda \frac{\partial \lambda}{\partial t}   \right)  \frac{\partial \delta q_{2} }{\partial t} 
 + \lambda^{2} \frac{ \partial^{2}\delta q_{2} }{\partial t^{2}}  \, . 
\end{aligned}
\end{equation}
We use the above expressions of the partial derivatives of $\delta \omega $ and $ \delta q_{2}$ from the equations \ref{A1} and \ref{A2} in the equation \ref{4.15}, and then we Fourier-transform both sides of the equation from time-space to frequency-space. For brevity, from now we designate the quantity $ \frac{\Delta \lambda}{R^{4}}   \frac{\partial}{\partial \overline{t}} \left(  \frac{R^{4}}{\Delta \lambda}  \right)  = \mathbb{F}(r,t) $ and the quantity  $ \frac{F}{R^{4}} \frac{\partial}{\partial \overline{t}} \Big( \frac{R^{4} }{F} \Big) = \mathscr{F}(r,t)$. Thus we obtain :
\begin{equation}  \label{4.20a}
\begin{aligned}
\frac{1}{2\pi} \int^{\infty}_{0}  \frac{F \lambda}{R^{4} \, Sin^{3} \theta}  \frac{\partial Q}{\partial \theta} \, e^{-i \sigma t}  dt  =
\\
 \frac{1}{2\pi} \int^{\infty}_{0}  F^{2} R \, Sin \, \theta  \Big( 1 - \frac{2 M_{H}}{R}   \Big)^{\frac{1}{2}} \Big(\frac{\lambda}{F} \Big)^{\frac{1}{2}}  \mathcal{R} \delta q_{2} \, e^{-i \sigma t}  dt
   \\
     - \frac{1}{2\pi} \int^{\infty}_{0}  \left\lbrace  a^{-1}\Big( 1 - \frac{M_{H}^{2}}{ 4 a^{2} r^{2} } \Big)^{-1}  \frac{\partial \delta \omega}{\partial r }    - \lambda \frac{\partial \delta q_{2}}{ \partial t}   \right\rbrace \mathscr{F} \, e^{-i \sigma t}  dt
     \\
 - \Big[   \frac{1}{2\pi} \int^{\infty}_{0} \lambda  \frac{\partial }{\partial t } \left\lbrace  a^{-1}\Big( 1 - \frac{M_{H}^{2}}{ 4 a^{2} r^{2} } \Big)^{-1}  \right\rbrace  \frac{\partial \delta \omega}{\partial r } \, e^{-i \sigma t}  dt    
 \\
 + \frac{1}{2\pi} \int^{\infty}_{0}  \lambda \left\lbrace   a^{-1}\Big( 1 - \frac{M_{H}^{2}}{ 4 a^{2} r^{2} } \Big)^{-1} \right\rbrace   \frac{\partial^{2} \delta \omega}{\partial r \partial t} \, e^{-i \sigma t}  dt 
 \\
  - \frac{1}{2\pi} \left\lbrace  \int^{\infty}_{0} \lambda  \frac{\partial \lambda}{\partial t }  \frac{\partial \delta q_{2}}{ \partial t} \, e^{-i \sigma t}  dt 
   +   \int^{\infty}_{0} \lambda^{2} \frac{\partial^{2} \delta q_{2} }{ \partial t^{2}} \, e^{-i \sigma t}  dt  \right\rbrace   \Big]    \,   .
\end{aligned}
\end{equation}
Similarly, after converting the partial derivatives of the perturbations w.r.t. $\overline{t} $ into that w.r.t. $t $,  then inserting those in the equation \ref{4.19}, and then Fourier-transforming both sides of the equation from time-space to frequency-space, we get :
\begin{equation} \label{4.21a}
\begin{aligned}
\frac{1}{2\pi} \int^{\infty}_{0} \frac{\Delta \lambda }{R^{4} \, Sin^{3} \theta } \frac{\partial}{\partial R} \left(  \lambda Q  \right) \, e^{-i \sigma t}  dt  
\\
= - \frac{1}{2\pi} \int^{\infty}_{0}  \lambda \Delta \, Sin \, \theta \, F \mathcal{R} \, \delta q_{3} \, e^{-i \sigma t}  dt
\\
+  \frac{1}{2\pi} \int^{\infty}_{0}  \left\lbrace  \lambda  \frac{\partial^{2} \delta \omega}{ \partial t \partial \theta }  
+   \mathbb{F} \frac{\partial \delta \omega}{\partial \theta} \right\rbrace  \, e^{-i \sigma t}  dt
\\
  + \frac{1}{2\pi} \int^{\infty}_{0}  \lambda \left\lbrace   - \frac{\partial \lambda}{\partial t} \frac{\partial \delta q_{3} }{ \partial t} 
   -\frac{ \partial \delta q_{3}}{\partial t} \mathbb{F}
  -   \lambda  \frac{\partial^{2} \delta q_{3}}{ \partial t^{2}} \right\rbrace  e^{-i \sigma t}  dt \, . 
\end{aligned}
\end{equation}
\vspace{0.5cm}\\
Now, applying the formula regarding Fourier-transformations and results of `Convolution-theorem' of Fourier-transformation for the product of two functions, which have been shown and derived in the APPENDIX-4 i.e. section \ref{Fourier-transform}, the equations \ref{4.20a} and \ref{4.21a} can be written as respectively :
\begin{widetext}
\begin{equation}    \label{4.20b}
\begin{aligned}
\left( \frac{F \lambda}{R^{4} \, Sin^{3} \theta} \right)^{\dagger} \frac{\partial Q^{\dagger}}{\partial \theta} =
  \left\lbrace  F^{2} R \, Sin \, \theta  \Big( 1 - \frac{2 M_{H}}{R}   \Big)^{\frac{1}{2}} \Big( \frac{\lambda}{F}  \Big)^{\frac{1}{2}} a^{-1}\Big( 1 - \frac{M_{H}^{2}}{ 4 a^{2} r^{2} } \Big)^{-1} \mathcal{R} \right\rbrace^{\dagger}  \delta q_{2}^{\dagger}  
   \\
     - \left\lbrace   a^{-1}\Big( 1 - \frac{M_{H}^{2}}{ 4 a^{2} r^{2} } \Big)^{-1} \mathscr{F}  \right\rbrace^{\dagger} \frac{\partial \delta \omega^{\dagger}}{\partial r } +  \left( \lambda \mathscr{F} \right)^{\dagger} ( i \sigma \delta q_{2} )^{\dagger}     
     \\
 - \Big[  \left\lbrace \lambda  \frac{\partial }{\partial t } \Big(  a \Big( 1 - \frac{M_{H}^{2}}{ 4 a^{2} r^{2} } \Big) \Big)^{-1}  \right\rbrace^{\dagger}  \frac{\partial \delta \omega^{\dagger}}{\partial r }    
 + \left\lbrace  \lambda  \Big(  a \Big( 1 - \frac{M_{H}^{2}}{ 4 a^{2} r^{2} } \Big) \Big)^{-1}  \right\rbrace^{\dagger}  i \sigma   \frac{\partial \delta \omega^{\dagger}}{\partial r } 
  -  \left\lbrace \lambda \frac{\partial \lambda}{\partial t } i \sigma 
   + \lambda^{2} (i \sigma )^{2} \right\rbrace^{\dagger}  \delta q_{2}^{\dagger}    \Big]    \,   ,
\end{aligned}
\end{equation}
\begin{center}
and
\end{center}
\begin{equation} \label{4.21b}
\begin{aligned}
\left(  \frac{\Delta \lambda^{2} }{R^{4} \, Sin^{3} \theta }  \right)^{\dagger} \left( \frac{\partial Q}{\partial R}  \right)^{\dagger}   +  \left(  \frac{\Delta \lambda}{R^{4} \, Sin^{3} \theta } \frac{\partial \lambda}{\partial R}   \right)^{\dagger}   Q^{\dagger}
\\
= - \delta q_{3}^{\dagger} \left( \lambda \Delta \, Sin \, \theta \, F \mathcal{R} \right)^{\dagger} + \left(    i \sigma  \lambda + \mathbb{F} \right)^{\dagger} \frac{\partial \delta \omega^{\dagger}}{\partial \theta} 
  + \left(   -i \sigma \lambda \frac{\partial \lambda}{\partial t} 
   - i \sigma  \lambda \mathbb{F} +
   \lambda^{2} \sigma^{2}  \right)^{\dagger} \delta q_{3}^{\dagger}   \, ,
\end{aligned}
\end{equation}
where the $ \dagger $-superscript over any bracket denotes the Fourier-transform of the quantity within the bracket from time-space to frequency-space. \\
Now, we have to eliminate $ \delta\omega^{\dagger} $ from these two equations \ref{4.20b} and \ref{4.21b}, so that we may obtain an equation describing the perturbations $\delta q_{2}^{\dagger} $ and $\delta q_{3}^{\dagger} $ only. For this, we start by partially differentiating these two equations w.r.t. $\theta $ and $r$ respectively, so that we may have the quantity $\frac{\partial^{2} \delta\omega^{\dagger}}{\partial \theta \partial  r} = \frac{\partial^{2} \delta\omega^{\dagger}}{\partial r \partial \theta }  $ in both the differentiated equations and we can replace that.
Hence, partially differentiating both sides of the equation \ref{4.20b} w.r.t. $ \theta $, we obtain the equation :
\begin{equation} \label{4.22}
\begin{aligned}
 \left(   \frac{F \lambda}{R^{4}}  \right)^{\dagger}  \frac{\partial}{\partial \theta} \Big( \frac{1}{Sin^{3} \theta}\frac{\partial Q^{\dagger}}{\partial \theta} \Big)   = 
  \left\lbrace  F^{2} R  \Big( 1 - \frac{2 M_{H}}{R} \Big)^{\frac{1}{2}} \Big( \frac{\lambda}{F}  \Big)^{\frac{1}{2}} \right\rbrace^{\dagger} \frac{\partial}{\partial \theta}(  \delta q_{2}  \mathcal{R} \, Sin \, \theta)^{\dagger} 
   \\
-  \left\lbrace  \lambda \frac{\partial}{\partial t } \Big(  a^{-1}\Big( 1 - \frac{M_{H}^{2}}{ 4 a^{2} r^{2} } \Big)^{-1} \Big) +  a^{-1}\Big( 1 - \frac{M_{H}^{2}}{ 4 a^{2} r^{2} } \Big)^{-1} \mathscr{F}  + 
i \sigma  \lambda a^{-1}\Big( 1 - \frac{M_{H}^{2}}{ 4 a^{2} r^{2} } \Big)^{-1}  \right\rbrace^{\dagger} \frac{\partial^{2} \delta \omega^{\dagger}}{\partial \theta \partial r}
\\
 +
 \left\lbrace - \lambda^{2}\sigma^{2} + i \sigma \left( \lambda \frac{\partial \lambda}{\partial t } 
 + \mathscr{F} \lambda  \right)  \right\rbrace^{\dagger} \frac{\partial \delta q_{2}^{\dagger}}{\partial \theta}   \, . 
\end{aligned}
\end{equation}    
Next, partially differentiating both sides of the equation \ref{4.21b} w.r.t. r, we obtain :
\begin{equation}\label{4.24}
\begin{aligned}
\frac{1}{Sin^{3} \theta} \frac{\partial }{\partial r } \left[  \left(  \frac{\Delta  \lambda }{R^{4}} \right)^{\dagger} \left\lbrace  \frac{\partial}{\partial R} \left( \lambda  Q \right)  \right\rbrace^{\dagger}  \right] 
= -Sin \, \theta \, \frac{\partial }{\partial r } \left\lbrace  \delta q_{3}^{\dagger} \left(  \lambda \Delta   F \mathcal{R}  \right)^{\dagger}  \right\rbrace
\\
+ \left(   i \sigma  \lambda + \mathbb{F} \right)^{\dagger} \frac{\partial^{2} \delta \omega^{\dagger}}{ \partial r  \partial \theta} 
+ \frac{\partial \delta \omega^{\dagger}}{\partial \theta } \frac{\partial }{\partial r }  \left(   i \sigma  \lambda + \mathbb{F}  \right)^{\dagger} 
  +  \left(  -i \sigma \lambda \frac{\partial \lambda}{\partial t}  -  i \sigma \lambda  \mathbb{F} +   \lambda^{2} \sigma^{2}  \right)^{\dagger} \frac{\partial \delta q_{3}^{\dagger} }{\partial r } 
   \\
 + \delta q_{3}^{\dagger} \frac{\partial}{\partial r } \left\lbrace   \lambda \left( -i \sigma \frac{\partial \lambda}{\partial t} 
  - i \sigma   \mathbb{F}   +   \lambda \sigma^{2} \right)  \right\rbrace^{\dagger}  \, .
\end{aligned}
\end{equation}
Again, for brevity, we designate the quantity $ a\Big( 1 - \frac{M_{H}^{2}}{ 4 a^{2} r^{2} } \Big)  \frac{\partial }{\partial \overline{ t}}  \left( a\Big( 1 - \frac{M_{H}^{2}}{ 4 a^{2} r^{2} } \Big) \right)^{-1} $ as $\mathcal{F} $ .
Now we have to substitute the quantity $ \frac{\partial^{2} \delta \omega^{\dagger}}{\partial \theta \partial r} $ from the equation \ref{4.22} in the equation \ref{4.24}.
But, the equation, which is obtained after substituting the expression of $\frac{\partial^{2} \delta \omega^{\dagger}}{\partial \theta \partial r} $ from the equation \ref{4.22} in the equation \ref{4.24}, would have a term containing $ \frac{\partial \delta \omega^{\dagger}}{\partial \theta } $. Therefore, in that equation, we have to again substitute the $ \frac{\partial \delta \omega^{\dagger}}{\partial \theta }  $ from the equation \ref{4.21b}, to make in completely in-terms of the perturbations $\delta q_{3}^{\dagger}$ and $\delta q_{2}^{\dagger} $. Doing these substitutions, we obtain :  
\begin{equation} \label{4.29}
\begin{aligned}
\frac{1}{Sin^{3} \theta} \frac{\partial }{\partial r } \left[   \left(  \frac{\Delta \lambda }{R^{4}} \right)^{\dagger}   \left\lbrace \frac{\partial}{\partial R} \left(   \lambda  Q \right) \right\rbrace^{\dagger} \right]  
=
\\
 -  \frac{ \left( i \sigma \lambda  + \mathbb{F} \right)^{\dagger} }{ \left\lbrace \left( i \sigma \lambda + \mathscr{F} + \mathcal{ F} \right) \left(  a\Big( 1 - \frac{M_{H}^{2}}{ 4 a^{2} r^{2} } \Big)\right)^{-1}  \right\rbrace^{\dagger} } \Big[  \left\lbrace \left(  \frac{F \lambda }{R^{4}}  \right)^{\dagger}  \frac{1}{Sin^{3} \theta}\Big( -3 Cot \, \theta \, \frac{\partial Q^{\dagger} }{\partial \theta } + \frac{\partial^{2} Q^{\dagger}  }{\partial \theta^{2}} \Big)  \right\rbrace
 \\
 -\left\lbrace  F^{2} R  \Big( 1 - \frac{2 M_{H}}{R} \Big)^{\frac{1}{2}} \Big( \frac{\lambda}{F} \Big)^{\frac{1}{2}} \right\rbrace^{\dagger}  \frac{\partial}{\partial \theta}(  \delta q_{2}  \mathcal{R} \, Sin \, \theta )^{\dagger}   \Big] - Sin \, \theta \, \frac{\partial }{\partial r } \left\lbrace  \delta q_{3}^{\dagger} \left( \lambda \Delta   F \mathcal{R} \right)^{\dagger}   \right\rbrace  
 \\
 + \left\lbrace  \lambda \left(  \lambda \sigma^{2} - i \sigma \frac{\partial \lambda}{\partial t} \right)   a\Big( 1 - \frac{M_{H}^{2}}{ 4 a^{2} r^{2} } \Big) \right\rbrace^{\dagger}  \left\lbrace \left(  \frac{\partial \delta q_{3}}{\partial R } \right)^{\dagger}  -   \frac{ \left( i \sigma \lambda  + \mathbb{F} \right)^{\dagger} }{  \left( i \sigma \lambda + \mathscr{F} + \mathcal{ F} \right)^{\dagger}}  \left( \frac{ \partial \delta q_{2} }{\partial \theta } \right)^{\dagger} \right\rbrace 
 \\
 +  \left(  i \sigma \lambda  \right)^{\dagger} \left[ \frac{   \left( i \sigma \lambda + \mathbb{F} \right)^{\dagger} }{\left\lbrace \left( i \sigma \lambda + \mathscr{F} + \mathcal{ F} \right) \left( a\Big( 1 - \frac{M_{H}^{2}}{ 4 a^{2} r^{2} } \Big) \right)^{-1} \right\rbrace^{\dagger} }   \left( \mathscr{F} \Big( \frac{ \partial \delta q_{2}}{\partial \theta } \Big) \right)^{\dagger}  - \left\lbrace \frac{ \mathbb{F} }{\left( a\Big( 1 - \frac{M_{H}^{2}}{ 4 a^{2} r^{2} } \Big) \right)^{-1}} \Big( \frac{\partial \delta q_{3}}{\partial R } \Big)\right\rbrace^{\dagger}  \right]  
 \\
 + \frac{ {\displaystyle \frac{\partial }{\partial r } \left( i \sigma  \lambda +  \mathbb{F} \right)^{\dagger} } }{ {\displaystyle \left( i \sigma  \lambda + \mathbb{F} \right)^{\dagger} }}  
\left\lbrace \left(  \frac{\Delta \lambda }{R^{4} \, Sin^{3} \theta } \right)^{\dagger} \left(  \frac{\partial}{\partial R} \left(  \lambda  Q  \right) \right)^{\dagger}  + 
 \delta q_{3}^{\dagger} \left( \lambda \Delta \, Sin \, \theta  F \mathcal{R}  \right)^{\dagger}  \right\rbrace   
 \\
 - {\displaystyle \left\lbrace  i \sigma \lambda \left( \frac{\partial \lambda}{\partial t}  + \mathbb{F} + i \sigma \lambda \right) \right\rbrace^{\dagger} }  
  \left[ - \frac{ {\displaystyle \frac{\partial }{\partial r} \left\lbrace  -i \sigma \lambda   \left( \frac{\partial \lambda}{\partial t}   +   \mathbb{F}  + i \sigma  \lambda \right)   \right\rbrace^{\dagger}  } }{ {\displaystyle \left\lbrace  -i \sigma \lambda \left(  \frac{\partial \lambda}{\partial t} + \mathbb{F} + i \sigma \lambda \right) \right\rbrace^{\dagger} } } + \frac{ {\displaystyle \frac{\partial }{\partial r } \left( i \sigma \lambda +  \mathbb{F} \right)^{\dagger} } }{ {\displaystyle \left( i \sigma  \lambda + \mathbb{F} \right)^{\dagger} }}  \right]  \delta q_{3}^{\dagger}  \,  . 
 \end{aligned}
\end{equation}
\end{widetext}
Some simplification is required before further proceeding with the equation \ref{4.29}. 
It is to be noted that some terms on the RHS of the equation \ref{4.29} contain the Ricci scalar $ \mathcal{R}$,  multiplied with other quantities, while in a Fourier-transformed form and the whole function where it appears is acted by partial derivative w.r.t. $r$ or $\theta $.   \\
We check the value of the Ricci scalar in this regard. 
Now, we show that at any arbitrary distance from a PBH, described by any arbitrary spherically symmetric and diagonal metric, the Ricci scalar vanishes when the cosmic fluid is radiation i.e. has the equation-of-state parameter $1/3$. This can be shown using the Einstein's equation in the following way. The Einstein's equation gives :
\begin{equation} \label{4.36}
R_{\mu \nu} - \frac{1}{2} \mathcal{R} g_{\mu\nu} = (8 \pi ) T_{\mu \nu}  \, . 
\end{equation} 
Contracting both sides of the above equation \ref{4.36} with $ g^{\mu \nu}$, we get 
\begin{equation} \label{4.37}
\mathcal{R} - \frac{4}{2} \mathcal{R} = - \mathcal{R}  =( 8 \pi )  T_{\mu \nu} g^{\mu \nu}  \, . 
\end{equation}
Now, we substitute the stress-energy tensor component for an imperfect fluid : 
\begin{equation} \label{4.38}
  T_{\mu \nu} =  (\rho + p)u_{\mu}u_{\nu} + p g_{\mu \nu} +  (\gamma_{\mu} u_{\nu}  + \gamma_{\nu} u_{\mu}) + \Pi_{\mu \nu}  \, ,
\end{equation}
in the RHS of the above equation \ref{4.37}. Here, $\gamma_{\mu} $ is the heat-flux vector and $\Pi_{\mu \nu} $ is the viscous-shear tensor for the concerned fluid ; while $u_{\mu} $ denotes the four-velocity of the fluid. Then, on the RHS of the equation \ref{4.37}, we get, after contracting $ T_{\mu \nu}$ with $g^{\mu \nu} $ :
\begin{center}
$T_{\mu \nu} g^{\mu \nu} = (\rho + p)u_{\mu}u_{\nu}g^{\mu \nu} +  p g_{\mu \nu} g^{\mu \nu} +$
 $(\gamma_{\mu} u_{\nu}  + \gamma_{\nu} u_{\mu}) g^{\mu \nu}  + \Pi_{\mu \nu}g^{\mu \nu} $
\end{center}
\begin{equation} \label{4.39}
\begin{aligned}
=  - (\rho + p) + 4p + (\gamma_{\mu} u^{\mu} + \gamma_{\nu} u^{\nu}) + \Pi_{\mu \nu}g^{\mu \nu} 
\\
= - \rho + 3 p + 2 \gamma_{\mu} u^{\mu} +  \Pi_{\mu \nu}g^{\mu \nu}  \, . 
\end{aligned}
\end{equation} 
In our case the concerned cosmic fluid is radiation, then $ p = \rho/3 $. Again, as the heat-flux vector is transverse to the world-lines, $\gamma_{\mu} u^{\mu} = 0  $. Using these finally we obtain, 
\begin{equation} \label{4.40}
T_{\mu \nu} g^{\mu \nu} =  \Pi_{\mu \nu}g^{\mu \nu}  \, . 
\end{equation}
Therefore, if we assume that the radiation is viscosity free or $ \Pi_{\mu \nu} = 0 $, then $ T_{\mu \nu} g^{\mu \nu} = 0 $. Then, the equation \ref{4.37} gives $ \mathcal{R} = 0 $.
Now, using the fact that, in the case of our interest the Ricci scalar vanishes everywhere around the black hole, the equation \ref{4.29} simplifies to :
\begin{widetext}
\begin{equation} \label{4.41}
\begin{aligned}
\frac{1}{Sin^{3} \theta} \frac{\partial }{\partial r } \left[   \left(  \frac{\Delta \lambda }{R^{4}} \right)^{\dagger}   \left\lbrace \frac{\partial}{\partial R} \left(   \lambda  Q \right) \right\rbrace^{\dagger} \right]  
=
\\
  -  \frac{ \left( i \sigma \lambda  + \mathbb{F} \right)^{\dagger} }{ \left\lbrace \left( i \sigma \lambda + \mathscr{F} + \mathcal{ F} \right) \left(  a\Big( 1 - \frac{M_{H}^{2}}{ 4 a^{2} r^{2} } \Big)\right)^{-1}  \right\rbrace^{\dagger} } \Big[  \left\lbrace \left(  \frac{F \lambda }{R^{4}}  \right)^{\dagger}  \frac{1}{Sin^{3} \theta}\Big( -3 Cot \, \theta \, \frac{\partial Q^{\dagger} }{\partial \theta } + \frac{\partial^{2} Q^{\dagger}  }{\partial \theta^{2}} \Big)  \right\rbrace
 \\
 + \left\lbrace  \lambda \left(  \lambda \sigma^{2} - i \sigma \frac{\partial \lambda}{\partial t} \right)   a\Big( 1 - \frac{M_{H}^{2}}{ 4 a^{2} r^{2} } \Big) \right\rbrace^{\dagger}  \left\lbrace \left(  \frac{\partial \delta q_{3}}{\partial R } \right)^{\dagger}  -   \frac{ \left( i \sigma \lambda  + \mathbb{F} \right)^{\dagger} }{  \left( i \sigma \lambda + \mathscr{F} + \mathcal{ F} \right)^{\dagger}}  \left( \frac{ \partial \delta q_{2} }{\partial \theta } \right)^{\dagger} \right\rbrace 
 \\
 +  \left\lbrace  i \sigma \lambda \mathbb{F} \left( a\Big( 1 - \frac{M_{H}^{2}}{ 4 a^{2} r^{2} } \Big) \right)  \right\rbrace^{\dagger} \left[ \frac{   \left( i \sigma \lambda + \mathbb{F} \right)^{\dagger} }{ \left( i \sigma \lambda + \mathscr{F} + \mathcal{ F} \right)^{\dagger} }   \left\lbrace \frac{\mathscr{F}}{\mathbb{F}} \Big( \frac{ \partial \delta q_{2}}{\partial \theta } \Big) \right\rbrace^{\dagger}  -  \Big( \frac{\partial \delta q_{3}}{\partial R } \Big)^{\dagger}  \right]  
 \\
 + \frac{ {\displaystyle \frac{\partial }{\partial r } \left( i \sigma  \lambda +  \mathbb{F} \right)^{\dagger} } }{ {\displaystyle \left( i \sigma  \lambda + \mathbb{F} \right)^{\dagger} }}
\left[ \left(  \frac{\Delta \lambda }{R^{4} \, Sin^{3} \theta } \right)^{\dagger} \left\lbrace   \frac{\partial}{\partial R} \left(  \lambda  Q \right) \right\rbrace^{\dagger}  \right] 
 \\
 - {\displaystyle \left\lbrace  i \sigma \lambda \left( \frac{\partial \lambda}{\partial t}  + \mathbb{F} + i \sigma \lambda \right) \right\rbrace^{\dagger} }  
  \left[ - \frac{ {\displaystyle \frac{\partial }{\partial r} \left\lbrace  i \sigma \lambda   \left( \frac{\partial \lambda}{\partial t}   +   \mathbb{F}  + i \sigma  \lambda \right)   \right\rbrace^{\dagger}  } }{ {\displaystyle \left\lbrace  i \sigma \lambda \left(  \frac{\partial \lambda}{\partial t} + \mathbb{F} + i \sigma \lambda \right) \right\rbrace^{\dagger} } } + \frac{ {\displaystyle \frac{\partial }{\partial r } \left( i \sigma \lambda +  \mathbb{F} \right)^{\dagger} } }{ {\displaystyle \left( i \sigma  \lambda + \mathbb{F} \right)^{\dagger} }}  \right]  \delta q_{3}^{\dagger}  \,  . 
 \end{aligned}
\end{equation}
\end{widetext}
At this stage, we note that the above equation \ref{4.41} can yet not be expressed completely in terms of the perturbation variable $Q^{\dagger}$. But, we see that after applying certain approximations, the above equation \ref{4.41} can be expressed w.r.t. $Q^{\dagger}$ only and subsequently the separation of variables technique can be applied to it. We describe this in the next section. 
\section{ The simplified equation after applying the approximations and separation of variables } \label{section-4}
\subsection{Separation of Radial and Angular Parts of the equation : }
After applying the approximations described in the APPENDIX-5 i.e. section \ref{Appendix-4}, including the approximation $ \frac{\left( i \sigma \lambda  + \mathscr{F} \right)^{\dagger}  }{ \left( i \sigma \lambda + \mathscr{F} + \mathcal{ F} \right)^{\dagger} }  \approx 1 $, 
 and expressing the quantities $Q_{23}^{\dagger}$ in terms of $Q^{\dagger}$, the equation \ref{4.41} can be written as :
\begin{equation} \label{8.3}
\begin{aligned}
 \left( \frac{\partial }{\partial R } \right) ^{\dagger} \left[ \left(   \frac{\Delta \lambda }{R^{4}} \right)^{\dagger}  \left\lbrace    \frac{\partial}{\partial R} \left( \lambda  Q \right)  \right\rbrace^{\dagger}  \right]  
 \\
=
 -    \left(  \frac{F \lambda }{R^{4}} \right)^{\dagger}   \Big( -3 Cot \, \theta \, \frac{\partial Q^{\dagger}}{\partial \theta } + \frac{\partial^{2} Q^{\dagger} }{\partial \theta^{2}} \Big)   
 \\
 +  \left\lbrace  \frac{\lambda}{\Delta} \left(  \lambda \sigma^{2} - i \sigma \frac{\partial \lambda}{\partial t} \right)  \right\rbrace^{\dagger} \left( -Q^{\dagger} \right) 
 + \left(  i \sigma \frac{\lambda  \mathscr{F}}{\Delta}  \right)^{\dagger}   Q^{\dagger} 
 \\ 
 +   \frac{ {\displaystyle \left( \frac{\partial }{\partial R } \right)^{\dagger}  \left( i \sigma  \lambda +  \mathbb{F} \right)^{\dagger} } }{ {\displaystyle \left( i \sigma \lambda + \mathbb{F} \right)^{\dagger} }} 
 \left( \frac{\Delta \lambda}{R^{4} } \right)^{\dagger}
\left\lbrace   \frac{\partial}{\partial R} \left(  \lambda  Q \right)  \right\rbrace^{\dagger}  \, .  
\end{aligned}
\end{equation}
Thereafter we express the quantity $ Q^{\dagger} $ as the multiplication of two parts $Q^{\dagger}_{R}(r, \sigma) $ and $Q^{\dagger}_{\theta} (\theta) $ as $Q^{\dagger} = Q^{\dagger}_{R}Q^{\dagger}_{\theta} $, where the part $Q^{\dagger}_{R} $, called the radial part, is a function of $r$ and $\sigma$ ; while the part $Q^{\dagger}_{\theta}$, called the angular part, is a function of $ \theta $. \footnote{ It is to be noted that, as we had inserted the fourier-transformed axial-perturbations into the equations in Section \ref{section-3}, from then their derivative w.r.t. isotropic time $ t $ is carried by multiplication with a factor of $i\sigma $ in fourier-space. The fourier-transformed axial-perturbations are independent on the isotropic time $t$, and is depedent on isotropic radial coordinate $r$, angular coordinate $\theta$ and frequency $\sigma$. 
}
 We now substitute $Q^{\dagger} = Q^{\dagger}_{R}Q^{\dagger}_{\theta} $ in the equation \ref{8.3}, to implement the procedure known as separation of variables technique and thus we obtain :
\begin{equation} \label{8.5}
\begin{aligned}
\frac{1}{Q^{\dagger}_{R}} \left(  \frac{R^{4}}{F\lambda}  \frac{\partial }{\partial R } \right)^{\dagger}  \left[ \left(   \frac{\Delta \lambda }{R^{4}} \right)^{\dagger}  \left\lbrace    \frac{\partial}{\partial R} \left( \lambda  Q_{R} \right)  \right\rbrace^{\dagger}  \right]  
\\
+ \left(   \frac{R^{4}}{F \Delta}  \right)^{\dagger} \left\lbrace  \left(  \lambda \sigma^{2} - i \sigma \frac{\partial \lambda}{\partial t}    \right)  - i \sigma \mathscr{F} \right\rbrace^{\dagger}
  \\
-  \frac{1}{Q^{\dagger}_{R}} \frac{ {\displaystyle \left( \frac{\partial }{\partial R }  \right)^{\dagger} \left( i \sigma  \lambda +  \mathbb{F} \right)^{\dagger} } }{ {\displaystyle \left( i \sigma  \lambda + \mathbb{F} \right)^{\dagger} }} 
 \left( \frac{\Delta }{F } \right)^{\dagger} \left\lbrace  \frac{\partial}{\partial R} \left(  \lambda  Q_{R} \right) \right\rbrace^{\dagger}
 \\
 = 
  -  \frac{1}{Q^{\dagger}_{\theta}}  \Big( -3 Cot \, \theta \, \frac{\partial Q^{\dagger}_{\theta}}{\partial \theta } + \frac{\partial^{2} Q^{\dagger}_{\theta} }{\partial \theta^{2}} \Big)     \, . 
\end{aligned}  
\end{equation}
Now, in the above equation \ref{8.5}, the LHS is a function of $r$ and $t$, 
 while the RHS is a function of $ \theta $. Therefore, we can say that in the above equation \ref{8.5}, the LHS = RHS = constant, which is independent of $r, \, t $ and $ \theta $. Let this constant be $\mathbb{K} $. Then, we can write two equations resulting from the equation \ref{8.5} as : 
\begin{equation} \label{8.6}
  - \frac{1}{Q^{\dagger}_{\theta}}   \Big(- 3 Cot \theta \frac{\partial Q^{\dagger}_{\theta}}{\partial \theta } + \frac{\partial^{2} Q^{\dagger}_{\theta}}{\partial \theta^{2}}    \Big) = \mathbb{K}
\end{equation}
\begin{center}
and
\end{center} 
\begin{equation}  \label{8.7}
\begin{aligned}
\frac{1}{Q^{\dagger}_{R}} \left(  \frac{R^{4}}{F\lambda}  \frac{\partial }{\partial R } \right)^{\dagger}  \left[ \left(   \frac{\Delta \lambda }{R^{4}} \right)^{\dagger}  \left\lbrace    \frac{\partial}{\partial R} \left( \lambda  Q_{R} \right)  \right\rbrace^{\dagger}  \right]  
\\
+ \left(   \frac{R^{4}}{F \Delta}  \right)^{\dagger} \left\lbrace  \left(  \lambda \sigma^{2} - i \sigma \frac{\partial \lambda}{\partial t}    \right)  - i \sigma \mathscr{F} \right\rbrace^{\dagger}
  \\
-  \frac{1}{Q^{\dagger}_{R}} \frac{ {\displaystyle \left( \frac{\partial }{\partial R }  \right)^{\dagger} \left( i \sigma  \lambda +  \mathbb{F} \right)^{\dagger} } }{ {\displaystyle \left( i \sigma  \lambda + \mathbb{F} \right)^{\dagger} }} 
 \left( \frac{\Delta }{F } \right)^{\dagger} \left\lbrace  \frac{\partial}{\partial R} \left(  \lambda  Q_{R} \right) \right\rbrace^{\dagger}
=  \mathbb{K}   \,  .
\end{aligned}
\end{equation}
So, we see that the equation satisfied by the angular part $Q^{\dagger}_{\theta} $ i.e. the equation \ref{8.6}, is same as that is obtained in case of the Schwarzschild metric, as expected. Hence, the angular part in this case too, like the Schwarzschild metric, can be taken as : $ Q^{\dagger}_{\theta } \propto \mathcal{C}^{-3/2}_{l+2} $ \cite{Chandrasekhar}; where $\mathcal{C}^{\nu}_{n} $ is the `Gegenbauer function'. This function $\mathcal{C}^{\nu}_{n} $ satisfies the equation : 
\begin{equation} \label{8.8}
\Big[  \frac{d}{d \theta} Sin^{2 \nu} \theta \frac{d}{d \theta} + n(n + 2 \nu) Sin^{2 \nu}\theta  \Big] \mathcal{C}^{\nu}_{n} = 0  \, . 
\end{equation}  
It may be noted that the `Gegenbauer function' $\mathcal{C}^{-3/2}_{l+2} $, associated with this case is related to the `Legendre Polynomial function' $ P_{l}(\theta)$ by the formulae : 
\begin{equation} \label{8.9}
\mathcal{C}^{-3/2}_{l+2}(\theta) = Sin^{3} \theta \frac{d}{d \theta} \Big(  \frac{1}{Sin \theta} \frac{d}{d \theta} P_{l}(\theta) \Big)  \, . 
\end{equation}
The angular part $ Q^{\dagger}_{\theta } $ being proportional to the $ \mathcal{C}^{-3/2}_{l+2} $, sets the constant $ \mathbb{K} $ and hence, the equation satisfied by the radial part $ Q^{\dagger}_{R}$  becomes : 
\begin{equation} \label{8.10}
\begin{aligned}
 \left(  \frac{R^{4}}{F\lambda}  \frac{\partial }{\partial R } \right)^{\dagger}  \left[ \left(   \frac{\Delta \lambda }{R^{4}} \right)^{\dagger}  \left\lbrace    \frac{\partial}{\partial R} \left( \lambda  Q_{R} \right)  \right\rbrace^{\dagger}  \right]  
 \\
+ \left(   \frac{R^{4}}{F \Delta}  \right)^{\dagger} \left\lbrace  \left(  \lambda \sigma^{2} - i \sigma \frac{\partial \lambda}{\partial t}    \right)  - i \sigma \mathscr{F} \right\rbrace^{\dagger} Q^{\dagger}_{R}
  \\
-   \frac{ {\displaystyle \left( \frac{\partial }{\partial R }  \right)^{\dagger} \left( i \sigma  \lambda +  \mathbb{F} \right)^{\dagger} } }{ {\displaystyle \left( i \sigma  \lambda + \mathbb{F} \right)^{\dagger} }} 
 \left( \frac{\Delta }{F } \right)^{\dagger} \left\lbrace  \frac{\partial}{\partial R} \left(  \lambda  Q_{R} \right) \right\rbrace^{\dagger}
 =
 \mathscr{M}^{2} Q^{\dagger}_{R} \,  ,
\end{aligned}
\end{equation}
where the quantity $\mathscr{M} = (l+2)(l-1) $, where $l$ is a positive integer $\geqslant 2 $, describes the angular dependence.
\par
 Let us see to which form the above equation \ref{8.10} reduces if we go from generalized McVittie metric to Schwarzschild metric of a non-rotating uncharged black hole of constant mass. We denote the radial coordinate and the constant mass of the Schwarzschild black hole as $r$ and $m$ respectively, while the perturbation-variable can be represented as simply $Q_{R}$, because its fourier-transformation is not required in the case of time-independent Schwarzschild metric. In case of Schwarzschild black hole of constant mass, the quantity $\lambda $ becomes 1 ($C$ becomes 0 and $F$ becomes 1). Also, the quantities $\mathbb{F}$ and $\mathscr{F}$ become zero, as these involve derivatives w.r.t. time. Hence, the equation \ref{8.10} reduces to :
\begin{equation} \label{8.10a}
\Delta \frac{d}{d r} \left( \frac{\Delta}{r^4} \frac{d Q_R}{d r} \right)  +  \sigma^2 Q_R - \mathscr{M}^2 \frac{\Delta}{r^4} Q_R = 0 \, , 
\end{equation}   
where the partial derivatives have been replaced by total derivatives w.r.t. r, as for Schwarzschild metric of constant mass the concerned quantities are time-independent. \\
The equation \ref{8.10a} is just another-form of the `Regge-Wheeler equation'. Therefore, we get the Regge-Wheeler equation from the equation \ref{8.10}, when generalized McVittie metric reduces to time-independent Schwarzschild metric. This in fact proves one facet of the correctness of the equation \ref{8.10}.  
\subsection{The horizons in the generalized McVittie metric : }
At this stage it is necessary to introduce the horizons of the black hole. The horizons of the generalized McVittie metric are given by the condition : 
\begin{equation} \label{h1}
\Big( 1 - \frac{C^{2}}{A^{2}} \Big) = 0 \Rightarrow A^{2} = C^{2} \, .     
\end{equation}
The expression of these horizons have been derived in the reference \cite{GaoChen} and these are given by : 
\begin{equation} \label{h2}
\begin{aligned}
R_{\pm} = \frac{1}{2H} \Big[  1 - M_{H}\Big( 1 + \frac{m}{2r}\Big) \frac{\dot{m}}{m}\pm 
\\
\sqrt{\left\lbrace  1 - M_{H} \Big(1+\frac{m}{2r} \Big) \frac{\dot{m}}{m} \right\rbrace^{2} - 8 m \dot{a}} \Big]  \, , 
\end{aligned}
\end{equation} 
where it is to be noted that as $R$ depends on $r$, the above expression of $R_{\pm}$ is implicit. Here, when the quantity inside the square-root i.e. $\left\lbrace  1 - M_{H} \Big(1+\frac{m}{2r} \Big) \frac{\dot{m}}{m} \right\rbrace^{2} - 8 m \dot{a} $ is positive, the horizons are physical. Then, the larger $R_{+}$ can be called the `Cosmic apparent horizon' and the smaller $R_{-}$ can be called the `Black hole apparent horizon'. It is quite clear that when the quantity inside the square-root on the RHS of equation \ref{h2} vanishes then these two horizons coincide and when this quantity is negative, there is no physical horizon, instead `naked singularity' comes there. The later two cases of coinciding horizons and `naked singularity' are not of interest in our case. \\
Therefore in this scenario, whenever, we have to deal with a description of the perturbations or any function of perturbations in this space-time, we have to confine that from the black hole apparent horizon to the cosmic apparent horizon.
\subsection{Change of variables and the boundary conditions :}
Now, we shall transform the radial coordinate from $ R$ to $ R_{\star}$, where the new coordinate $R_{\star} $ is defined as \footnote{It is to be noted that $R_{\star} $ is not a `tortoise-coordinate' in this case.} : 
\begin{equation} \label{8.11}
\frac{\partial}{\partial R_{\star}} \equiv  \left( \frac{\Delta }{R^{2}  } \frac{\partial}{\partial R} \right)^{\dagger}   \, . 
\end{equation}
Again, for convenience, we define a new variable $ \tilde{Q}_{R} = (\lambda Q_{R})^{\dagger} $. The interesting issue about this variable is that as $\Big( 1 - \frac{C^{2}}{A^{2}}  \Big) = 0 $ at the horizons $R_{\pm} $, this variable vanishes at both the horizons, except for zero frequency modes. Also, it should be mentioned that although the boundary conditions for $Q^{\dagger}_{R}$ has to be purely ingoing at the black hole apparent horizon $R_{-}$ and purely outgoing at the cosmic apparent horizon $R_{+}$, that does not affect the vanishing of the new variable $\tilde{Q}_{R} $ at both the horizons. \\ 
Using the new radial coordinate $ R_{\star} $ and the new variable $ \tilde{Q}_{R}$, putting $\mathscr{F}$ in place of $\mathbb{F}$ (as $ \mathbb{F} \approx \mathscr{F} $, the approximation which we have already applied), the above equation \ref{8.10} can be written as :
\begin{equation} \label{8.14}
\begin{aligned} 
\frac{\partial^{2} \tilde{Q}_{R}}{\partial R_{\star}^{2}}  + 
\left[  \frac{ {\displaystyle \frac{\partial }{\partial R_{\star}} \left(  \frac{\lambda}{R^{2}} \right)^{\dagger} }}{ {\displaystyle \left( \frac{\lambda}{R^{2}} \right)^{\dagger} } }    - \frac{ {\displaystyle \frac{\partial }{\partial R_\star}  \left( i \sigma  \lambda +  \mathscr{F} \right)^{\dagger} } }{ {\displaystyle \left( i \sigma  \lambda + \mathscr{F} \right)^{\dagger} }}      \right] 
 \frac{\partial \tilde{Q}_{R}}{\partial R_{\star}}   
\\
+ \left\lbrace   \sigma^{2} - i \sigma \left( \frac{1}{\lambda} {\displaystyle \frac{\partial \lambda}{\partial t}  }   +   \frac{\mathscr{F}}{ {\displaystyle \lambda } } \right)^{\dagger} \right\rbrace   \tilde{Q}_{R}  = \mathscr{M}^{2} \left\lbrace \frac{\Delta }{R^{4}} \left( \frac{\lambda}{F}\right)^{-1} \right\rbrace ^{\dagger} \tilde{Q}^{\dagger}_{R}   \, . 
\end{aligned} 
\end{equation}
Sometimes it may be necessary to make the equation \ref{8.14} free of the single-derivative term. So, we briefly describe the transformation of the equation \ref{8.14} into the form containing only second-order derivative term. 
The process of elimination of first-order derivative term, from a general second-order differential-equation, is quite well-known. Let us first write the equation \ref{8.14} with the following notations for brevity : 
\begin{equation} \label{8.15}
\frac{\partial^{2} \tilde{Q}_{R}}{\partial R_{\star}^{2}}  +  \xi(r,\sigma) \frac{\partial \tilde{Q}_{R}}{\partial R_{\star}}  + \zeta(r,\sigma) \tilde{Q}_{R}  =  0   \, , 
\end{equation}   
where the quantities $ \xi(r,\sigma) $ and $  \zeta(r,\sigma) $ stand for respectively : 
\begin{equation} \label{8.16a}
\begin{aligned}
\xi(r,\sigma) =
  \left[  \frac{ {\displaystyle \frac{\partial }{\partial R_{\star}} \left(  \frac{\lambda}{R^{2}} \right)^{\dagger} }}{ {\displaystyle \left( \frac{\lambda}{R^{2}} \right)^{\dagger} } }    - \frac{ {\displaystyle \frac{\partial }{\partial R_\star}  \left( i \sigma  \lambda +  \mathscr{F} \right)^{\dagger} } }{ {\displaystyle \left( i \sigma  \lambda + \mathscr{F} \right)^{\dagger} }}      \right]    \, , 
\end{aligned} 
\end{equation} 
\begin{center}
and
\end{center}
\begin{equation} \label{8.16b}
\begin{aligned} 
\zeta(r,\sigma) = 
 \left\lbrace   \sigma^{2} - i \sigma \left( \frac{1}{\lambda} {\displaystyle \frac{\partial \lambda}{\partial t}  }   +   \frac{\mathscr{F}}{ {\displaystyle \lambda } } \right)^{\dagger} \right\rbrace   
 -  \mathscr{M}^{2} \left\lbrace \frac{\Delta }{R^{4}} \left( \frac{\lambda}{F} \right)^{-1}  \right\rbrace^{\dagger}  \, . 
\end{aligned}
\end{equation}
Now, we use the substitution : 
\begin{equation} \label{8.17}
\Psi_{R} = exp.\left\lbrace \int^{R_{\star} } \xi \, dR' \right\rbrace \tilde{Q}_{R}(R_{\star})  
\end{equation}
in the equation \ref{8.15}. 
 With this substitution, after some calculations the equation \ref{8.15} is transformed into the form : 
\begin{equation} \label{8.18}
\frac{\partial^{2} \Psi_{R}}{\partial R_{\star}^{2}} +   \left\lbrace  \zeta - \frac{1}{2} \Big(\frac{\partial \xi}{\partial R_{\star}}  \Big)- \frac{1}{4} \xi^{2} \right\rbrace \Psi_{R} =  0 \, . 
\end{equation}  
The above equation \ref{8.18} may be called the \textsl{equivalent-counterpart of the `Regge-Wheeler equation' for generalized McVittie metric } : the equation governing the `Axial perturbations' in the space-time of a Schwarzschild black hole embedded in FLRW-Universe, described by the generalized McVittie metric.
\begin{widetext}
The equation \ref{8.14} has now been transformed into the equation \ref{8.18}, which is a Schr\"{o}dinger-like form and we rewrite this in the following style :   
\begin{equation} \label{8.19}
\begin{aligned}
\frac{\partial^{2} \Psi_{R}}{\partial R_{\star}^{2}} + \sigma^2  \Psi_{R} =
 \left[   i \sigma \left( \frac{1}{\lambda} {\displaystyle \frac{\partial \lambda}{\partial t}  }   +   \frac{\mathscr{F}}{ {\displaystyle \lambda } } \right)^{\dagger}   
 +  \mathscr{M}^{2} \left\lbrace \frac{\Delta }{R^{4}} \left( \frac{\lambda}{F} \right)^{-1}  \right\rbrace^{\dagger}
 + \left\lbrace  \frac{1}{2} \Big(\frac{\partial \xi}{\partial R_{\star}}  \Big)+  \frac{1}{4} \xi^{2} \right\rbrace \right]  \Psi_{R} \, .
\end{aligned} 
\end{equation}
\end{widetext}
\subsection{Some Physical interpretations from the Potential :}
From the equation \ref{8.19} in Schr\"{o}dinger-like form, we can say that the effective potential for $\Psi_{R}$ is given by : 
\begin{equation} \label{9.1}
\begin{aligned}
V = \Big[   i \sigma \left( \frac{1}{\lambda} {\displaystyle \frac{\partial \lambda}{\partial t}  }   +   \frac{\mathscr{F}}{ {\displaystyle \lambda } } \right)^{\dagger}   
 +  \mathscr{M}^{2} \left\lbrace \frac{\Delta }{R^{4}} \left( \frac{\lambda}{F} \right)^{-1}  \right\rbrace^{\dagger}
 \\
 + \left\lbrace  \frac{1}{2} \Big(\frac{\partial \xi}{\partial R_{\star}}  \Big)+  \frac{1}{4} \xi^{2} \right\rbrace \Big] \, , 
\end{aligned}
\end{equation}
which is a function of both the radial coordinate and frequency. \\
Some important characteristics of this potential can be noted. Before going into details, we need to describe certain properties of the quantities $ R, \, \lambda, \, \Delta, \,  F $ etc., in context of their transformation due to sign-reversal of time coordinate. It is to be noted that some of these quantities and their certain functions are even w.r.t. time coordinates $t$ and $\overline{t}$, which can be examined by investigating their transformation under a sign-reversal of these time coordinates (i.e. $t \rightarrow -t $, $\overline{t} \rightarrow -\overline{t}$). 
\\
First of all, we note the property of the quantity $C = HR + \dot{m}a \sqrt{\tilde{r}/r }$. It is evident that $R, \, m,$ and $a$ must be even w.r.t. $t$. This is because radial distances $R, \, \tilde{r}$ or $r$ and mass $m$ can not be negative. A negative scale factor $a$ is also unphysical according to the FLRW-metric. So, the only quantities that change sign under the transformation $t \rightarrow -t $ are $H$ and $\dot{m}$, as these contain the derivative w.r.t. $t$. So,  under the transformation $t \rightarrow -t $, $C$ changes sign viz. $C \rightarrow -C $, or, $C$ is an odd-function of $t$. Again, the quantity $A = 1 - 2 M_H / R $ clearly does not change sign under the transformation $t \rightarrow -t $, as it is made up of the parameters $a, \, m, $ and $r$. Similarly, the quantity $\Delta = R^2 - 2 M_H R $ is also even w.r.t, $t$.  \\
So, the quantity $\left( 1 - C^2/A^2 \right)$ is even under sign-reversal of $t$ (due to the squared appearance of the quantity $C$). 
\\
We note the equation \ref{3.1} governing the integrating factor $F$, which is given in Appendix-3 i.e. section \ref{RelationsforGMcM}. The quantity $\beta$ on the RHS of the equation \ref{3.1}, is clearly odd under the sign-reversal of $t$, as its denominator $A^2 - C^2$ is even and numerator $ C $ is odd under the transformation $t \rightarrow -t$. Now, if we denote the solution of this equation as $F'$ for the sign-reversed case of $ t \rightarrow -t $, then it is simple to check that the equation for $F'$ and $F$ will be same. Hence, $ F $ should be unchanged under the transformation $t \rightarrow -t$ or, $F$ is an even function under the sign-reversal of $t$. 
As $ F $ and $\left( 1 - C^2/A^2 \right)$ both are even under the transformation  $t \rightarrow -t$, the quantity $\lambda = F \left( 1 - C^2/A^2 \right) $ is also even under sign-reversal of $t$.\\
Another fact is pertinent to mention here, although it is quite clear from the above discussion, for the line-element in `Nolan-gauge' (Schwarzschild-like coordinates) in \ref{1.10}, from the definition of $\overline{t}$, it follows that for a sign-reversal in $t$ implies a similar transformation in $\overline{t}$ i.e. $t \rightarrow -t \Rightarrow \overline{t} \rightarrow - \overline{t}$. \\
Again, the quantity $ \mathscr{F} = \frac{F}{R^4} \frac{\partial}{\partial \overline{t}} \frac{R^4}{F} $ is an odd function under sign-reversal of time $t$ or $\overline{t}$.   
\par 
Therefore from the above analysis, we see that in the potential $V$, the quantity  $\left( {\displaystyle \frac{1}{\lambda} } {\displaystyle \frac{\partial \lambda}{\partial t}  }   +   \frac{ {\displaystyle \mathscr{F} }}{ {\displaystyle \lambda } } \right) $ is an odd function of time $t$ and the quantity $\left\lbrace \frac{\Delta }{R^{4}} \left( \frac{\lambda}{F} \right)^{-1}  \right\rbrace $ is an even function of time $t$. So, the fourier-transforms of the quantities $ \left( {\displaystyle \frac{1}{\lambda} } {\displaystyle \frac{\partial \lambda}{\partial t}  }   +   \frac{ {\displaystyle \mathscr{F} }}{ {\displaystyle \lambda } } \right) $ and $\left\lbrace \frac{\Delta }{R^{4}} \left( \frac{\lambda}{F} \right)^{-1}  \right\rbrace $ are completely an imaginary quantity and completely a real quantity respectively.  
In the quantity $ \xi $, $ \lambda / R^2 $ is an even function of time $t$ and hence the first part $ { \left( \frac{\lambda}{R^{2}} \right)^{\dagger} }^{-1} \frac{\partial }{\partial R_{\star}} \left(  \frac{\lambda}{R^{2}} \right)^{\dagger}  $ within $\xi$ is completely a real quantity. In the second part of $\xi$, $ \mathscr{F}$ is an odd function and $\lambda $ is an even function of time $t$. 
\par
Hence, it is quite clear that if the frequency $\sigma $ is a real quantity, then the overall potential $V$ given in equation \ref{9.1} is a completely real quantity. 
\footnote{If $\sigma $ is real, then $i \sigma \lambda^{\dagger} $ and $\mathscr{F}^{\dagger} $
are both completely imaginary quantities and as a result $ \frac{ \frac{\partial }{\partial R_\star}  \left( i \sigma  \lambda +  \mathscr{F} \right)^{\dagger} }{ \left( i \sigma  \lambda + \mathscr{F} \right)^{\dagger} } $ is a completely a real quantity. }  
This indicates the fact that the potential would be complex, only if the frequency $\sigma$ is complex quantity. Or, in other way it can be said that the imaginary part of the frequency $\sigma$ is responsible for the imaginary part of the    potential. As a consequence, the stability of the system i.e. whether certain modes of axial perturbations are   unstable i.e. if $\sigma_{I} < 0 $ (viz. when the imaginary part of the frequency $\sigma_{I}$ is negative) \footnote{ Substituting $\sigma = \sigma_{R} + i \sigma_{I} $ in the $e^{ i \sigma t} $, one gets $e^{ i \sigma t} = e^{ i (\sigma_{R} + i \sigma_{I}) t} = e^{ i \sigma_{R} t} e^{- \sigma_{I} t}  $ .} can be examined by styding the imaginary part of the potential.  
\section{Conclusion and Discussion :} \label{section-6}
In this work, we have derived the equation governing the axial perturbations in the generalized McVittie metric, which 
can well describe the space-time around a non-rotating uncharged PBH, created in the early radiation-dominated Universe, where the effect of expansion of the Universe on the local space-time of the PBH was significant due to very high value of Hubble-parameter at that time and the PBHs were continuously changing masses due to spherical accretion of the surrounding high-density radiation.\\ 
In the process of deriving the equation in desired form, we have done Fourier-transformation of the overall perturbation equations from time-space to frequency-space and then applied several outcomes of the `Convolution-theorem' for Fourier transform of product of two functions. This procedure of Fourier-transforming the overall perturbation equations, whence not only the perturbation variables but also the unperturbed parameters are time-varying, and then application of the `Convolution-theorem', to get the equation in a suitable form, should be applicable for deriving equations governing perturbations in case of any time-dependent metric. It may be noted that generally in case of perturbations in any time-dependent black hole metric, either the temporal-part is separated from the spatial part or suitably chosen ansatz is applied for getting the equation in preferable format (for example see the reference \cite{Lin_et_al}.).  
\\
From the equation governing the axial perturbations in Schr\"{o}dinger-like form, which is free of first-order derivative term, we see that the potential is a complex-quantity and we have also explained that its imaginary part originates due to the imaginary part of the frequency. So, by examining the imaginary part of the potential we can have important information about the imaginary part of the frequency for any certain mode. 
\\
  We know that for some of the usual black hole metrics e.g. Schwarzschild, Schwarzschild-Anti-De Sitter etc., the unstable modes do not exist i.e. perturbations, exponentially growing with time, are not practically possible. As the generalized McVittie metric is physically the space-time around a Schwarzschild black hole of varying mass, embedded in an expanding FLRW-Universe, if unstable modes are found to exist, then it can be interpreted that the effects of expansion of the background FLRW-Universe or the time-variation of mass of the black hole are making the existence of the unstable modes possible. \\
For any unstable mode, as the corresponding axial-perturbation grows exponentially with time, at a certain stage the linear perturbation theory breaks down if the growing perturbation becomes sufficiently large so that it can not be treated by linear perturbation theory. In other way to say, these unstable modes indicate the possibility of non-linear effects in these perturbations. In some earlier works, the possibility of non-linear instabilities have been proposed in fast spinning black holes, with a similarity of turbulence in hydrodynamics \cite{Yang_et_al}. But till date, there is hardly any highly-energetic astrophysical or cosmological phenomena studied in numerical general relativity (e.g. the merging of two black holes in a binary), which can generate observationally important non-linear effects.    
 So in case of the generalized McVittie metric, it will be interesting to investigate if the unstable modes exists, then whether those can give rise to significant non-linear effects. If there exists significant non-linear instability, then that might even leave imprint on the stochastic gravitational wave background produced due to the vibration of perturbed PBHs in the early Universe.  \\
In some cases, non-zero stress-energy surrounding any black hole, which is then called `dirty black hole', can affect the linear stability of that black hole \cite{Boonserm_et_al}. This stress-energy can be even due a shell of matter or a planet. So, this indicates a concordance with the case of the generalized McVittie metric, where the mass of the black hole is time-varying due to spherical accretion of the surrounding radiation in the early radiation-dominated era. Here, the surrounding radiation should provide the non-zero stress-energy, which can affect the black hole's linear stability. Although in practical case, there are thought to be many ways to perturb those PBHs, as we have argued in the introduction (section \ref{section-1}).  
\\
The equation derived by us, is the preliminary step for investigating the conditions of stability or instability of  non-rotating uncharged PBHs of changing masses, described by the generalized McVittie metric, in the early radiation-dominated Universe. Though the similar counterpart for polar-perturbations is required too. 
In future we shall try to investigate the existence of the instability in this case, in terms of various parameters, more specifically.         
\section{APPENDIX 1 : Clarification about some dimensional issues }\label{DimensionalIssue}
We have to be clear about the presence of the two fundamental constants G (Universal Gravitational constant) and c (Speed of light in vacuum) in all the expressions, which are generally omitted according to the natural units' convention of taking G,c as 1 (unity). The convention of natural units is okay for purely analytical i.e. non-numerical calculations, but this is not suitable as we need to get exact numerical orders of several quantities.\\
For the second part of the quantity $C$ i.e. $ \dot{m}a \sqrt{\frac{\tilde{r}}{r}} $, we first evaluate $ \dot{m} $ in terms of the Hawking-Hayward Quasilocal mass $M_{H}$, which is related with the former as $M_{H}(t)= m(t)a(t) $ : 
\begin{center}
${\displaystyle \dot{m} = \frac{\dot{M}_{H}}{a} - \Big( \frac{\dot{a}}{a^{2}} \Big) M_{H}  } \, , $
\end{center} 
\begin{equation} \label{5.1}
\Rightarrow \dot{m} a = \dot{M}_{H} - H M_{H} \,  . 
\end{equation}
To estimate an approximate order of the first term on the RHS in the equation \ref{5.1} i.e. $ \dot{M}_{H}$, we use its expression derived in the reference \cite{Faraoni}. 
This gives the time-rate of change of the Hawking-Hayward Quasi-local mass in terms of cosmic-fluid density at a finite radial distance from the black hole (in isotropic coordinates) :
\begin{equation} \label{5.2a}
\dot{M}_{H} = -\frac{G}{2} a \mathcal{B}^{2} \sqrt{ 1 + a^{2} \mathcal{A}^{4}{u^{r}}^{2} } (P(r) + \rho(r)) \mathscr{A} u^{r}   \, , 
\end{equation}    
where the concerned quantities $\mathcal{A}$ and $\mathcal{B}$ have already been defined earlier. $u^{r} $ is the contra-variant radial component of the four-velocity of cosmic fluid getting spherically accreted by the black hole, $\mathscr{A}= 4\pi \mathcal{A}^{4} a^{2}r^{2} $ is the area of the spherical surface of isotropic radius $r $, $\rho(r)$ and $ P(r) $ are respectively the density and pressure of the cosmic fluid at that isotropic radial distance $r $.
 \\
 But, it is easy to verify that this expression of $\dot{M}_{H} $, when converted to the Schwarzschild-like coordinate in Nolan gauge, it gives :
 \begin{equation} \label{5.2b}
 \dot{M}_{H} = 4 \pi c R_{-}^{2} (1+ w) \rho  \, , 
 \end{equation}
 which is simply the time rate of accretion of cosmic-fluid into the black hole, through the apparent black hole horizon $R_{-} $. Here, $w $ is the equation-of-state parameter of the cosmic-fluid and as it is radiation in our case, $w = 1/3 $. \\
From the metric given in equation \ref{1.10}, it is clear that the quantity $C = HR + \dot{m} a\sqrt{\frac{\tilde{r}}{r}} $, written in the convention of natural units, must be dimensionless. In this context it is also to be noted that the corresponding ratio $M_{H}/R$ in the metric \ref{1.10}, is actually $GM_{H}/c^{2}R $ i.e. dimensionless, as the quantity $G/c^{2} $ remain omitted when we use the convention of natural units, as has been stated already. Hence, both the quantities $ HR $ and $  \dot{m} a\sqrt{\frac{\tilde{r}}{r}}  $ must be actually dimensionless. We here use the notations viz. [L],[M] and [T] for the dimensions of length, mass and time respectively. \\ 
 In the second term of C, in $\dot{m} $, including the $G/c^{2} $ that occurs with the mass $ m $ to make it a length-scale, we get $G\dot{m}/c^{2} $, which has the dimension of length/time : $[LT^{-1}]$ (Dimension of G : $[G] \equiv [L^{3} T^{-2} M^{-1} ]  $). So, although it seems that $[C] \equiv [HR] \equiv [\dot{m} a\sqrt{\frac{\tilde{r}}{r}} ]  \equiv [LT^{-1}] $, it is not the actual dimension, as $C$ must be dimensionless. So, for being dimensionless there must be a $c^{-1}$ multiplied with these and which is evident as we shall show later that $C = \frac{\partial R}{\partial t } \equiv  \frac{\partial R}{c\partial t }  $. 
Therefore, without erasing G and c the actual expression of $C$ is :
\begin{equation} \label{5.3}
 C(t,R) = \frac{HR}{c} + \frac{G\dot{m}}{c^{3}} a\sqrt{\frac{\tilde{r}}{r}} 
\end{equation}
It can be easily verified that C is dimensionless by substituting the dimensions of corresponding quantities.  \\
The confusion for the expression of $\dot{M}_{H} $ given in equation \ref{5.2a} is deeper if we do not write the omitted G, c in proper places because the author in reference \cite{Faraoni} has kept the `G' in Einstein's equations there, while has omitted the `c's there and also omitted those `G',`c's present with the masses in the ratios $M_{H}/r \equiv GM_{H}/c^{2}r $ in metric coefficients. It can be easily checked that for $\dot{M}_{H} $ to be dimensionless there must be $Gc^{-3}$ with it.
  This can be checked by substituting the dimensions of the corresponding quantities.
    Thus, keeping the `G' and `c's in proper places the actual expressions are :
\begin{equation} \label{5.4}
{\displaystyle   \frac{G\dot{M}_{H}}{c^{3}} = -\frac{G}{2c^{3}} a B^{2} \sqrt{ 1 + a^{2}A^{4} \Big(\frac{u^{r}}{c}\Big)^{2} } (P(r) + \rho(r)) \mathscr{A} u^{r}     \,  ,} 
\end{equation}
Or, canceling $Gc^{-3} $ from both sides,
\begin{equation} \label{5.5}
  \dot{M}_{H} = -\frac{1}{2} a B^{2} \sqrt{ 1 + a^{2}A^{4} \Big(\frac{u^{r}}{c}\Big)^{2} } (P(r) + \rho(r)) \mathscr{A} u^{r}       \, . 
\end{equation}
\section{APPENDIX 2 : Basic equations describing the Axial perturbations } \label{BasicPertEqn}
Here, the origin of the equations governing the perturbations on a metric given by equation \ref{4.1}, has been described.
In this case, the corresponding unperturbed components of the Ricci tensors are :
\begin{equation}
\begin{aligned} \label{6.2}
R_{12} = \frac{1}{2} e^{-2\psi} (e^{-2 \nu} e^{-2 \mu_{3}})^{1/2} [ ((e^{3\psi + \nu -(\mu_{2} + \mu_{3})})
\\
 \mathcal{Q}_{32})_{,3} -  ((e^{3\psi - \nu -\mu_{2} + \mu_{3}}) \mathcal{Q}_{02})_{,0} ]  \, , 
\end{aligned}
\end{equation}
and 
\begin{equation}  \label{6.3}
\begin{aligned}
R_{13} = \frac{1}{2}e^{-2\psi} (e^{-2 \nu} e^{-2 \mu_{3}})^{1/2} [ ((e^{3\psi + \nu -(\mu_{2} + \mu_{3})}) 
\\
\mathcal{Q}_{23})_{,2} -  ((e^{3\psi - \nu +\mu_{2} - \mu_{3})}) \mathcal{Q}_{03})_{,0} ] \, , 
\end{aligned}
\end{equation}
where $ \mathcal{Q}_{AB} = q_{A, B }- q_{B, A }  $ and $ \mathcal{Q}_{A0} = q_{A, 0 }- \omega_{,A } $ are defined similarly as of $Q_{AB} $ and $Q_{A0} $, but they contain the unperturbed values of the quantities $q_{2}, \, q_{3} $ and $ \omega$ instead of their linear perturbations. \\
As for the metric given in equation \ref{4.1}, in our case, $ q_{2} = q_{3}= \omega = 0$, therefore $\mathcal{Q}_{AB}= \mathcal{Q}_{A0} = 0 $. Hence, the unperturbed Ricci tensor components $ R_{12}= R_{\phi R} $ and $ R_{13}= R_{\phi \theta} $ are zero (0).
 \\
The origin of the equations governing the perturbations is from Einstein's equations for those components. The Einstein's equation for these components, taken to first order axial perturbations, is given by :
\begin{equation} \label{6.5}
\begin{aligned}
R_{ij}+ \delta_{\omega,q_{2},q_{3}}  R_{ij} - \frac{1}{2} (g_{ij} +  \mathcal{R} \delta_{\omega,q_{2},q_{3}} g_{ij} + g_{ij}\delta_{\omega,q_{2},q_{3}}  \mathcal{R}) =
\\
(8 \pi)  (T_{ij}+ \delta_{\omega,q_{2},q_{3}} T_{ij}) \, ,
\end{aligned}
\end{equation}
where $i, j$ are indices denoting spatial-coordinates. 
(For avoiding confusion with the radial coordinate R, we denote the Ricci scalar by $ \mathcal{R}$.)\\
Subtracting the unperturbed Einstein's equation from the above equation \ref{6.5} with the linear perturbation, we obtain :
\begin{equation} \label{6.6}
\delta_{\omega,q_{2},q_{3}}  R_{ij} - \frac{1}{2} (g_{ij} \delta_{\omega,q_{2},q_{3}}  \mathcal{R} + \delta_{\omega,q_{2},q_{3}} g_{ij} \mathcal{R}) 
=(8 \pi)  \delta_{\omega,q_{2},q_{3}} T_{ij} \,  .
\end{equation}
As there is no cross components in the metric, hence for this case i.e. i=1 and j=2,3 ; $g_{12}=g_{13}=0  $, which reduces the above equation to :
\begin{equation} \label{6.7}
\delta_{\omega,q_{2},q_{3}}  R_{ij} - \frac{1}{2}  (\delta_{\omega,q_{2},q_{3}} g_{ij}) \mathcal{R}
= (8 \pi)  \delta_{\omega,q_{2},q_{3}} T_{ij}  \,          .
\end{equation}
We have already shown in the section \ref{section-3} that if the cosmic-fluid is radiation having equation-of-state parameter $w = 1/3$, then the associated Ricci-scalar vanishes. Hence, the equation \ref{6.7} further reduces to :
\begin{equation}  \label{6.8}
\delta_{\omega,q_{2},q_{3}}  R_{ij}  =  (8 \pi)  \delta_{\omega,q_{2},q_{3}} T_{ij}    \, .
\end{equation}  
While it can be shown that in a flat FLRW-Universe, for the part of the energy-momentum tensor belonging to a perfect fluid, the concerned  components of the linear perturbations of energy-momentum tensor of the cosmic-fluid are zero : $ \delta T_{ij \, p}= 0 $, where $i \neq j $ and the suffix `p' in the energy-momentum tensor component represents it is due to the `perfect' part of the fluid. But, in this case of generalized McVittie metric, a perfect fluid can not describe the surrounding cosmic-fluid. As it has been already shown and explained in the reference~\cite{Faraoni} that a single perfect cosmic-fluid can not describe a physical solution of a spherically symmetric black hole embedded in an expanding FLRW-Universe. We need at least one imperfectness parameter in it to describe the solution physically. As was chosen by the authors in reference \cite{Faraoni}, we also choose this imperfectness parameter to be the heat-flux vector $\gamma_{\mu}$. Only one component of the heat-flux vector suffices in this case and we can take it to be the radial component, in accordance with the radial mass-flow into the accreting black hole. So, due to the heat-flux vector there would be an additional imperfect part in the energy-momentum tensor of the cosmic-fluid  i.e. radiation in this case, which is $T_{ij\, Ip} = \gamma_{i}u_{j} + \gamma_{j}u_{i} $ (where `Ip' represents `imperfect'). So, the perturbation to this is given by (we write $\delta $ in place of $ \delta_{\omega,q_{2},q_{3}}  $ for brevity):
\begin{equation} \label{6.9}
 \delta T_{ij\, Ip} = u_{j} \delta \gamma_{i} + \gamma_{i} \delta u_{j} + u_{i} \delta \gamma_{j} +\gamma_{j} \delta u_{i} \, . 
 \end{equation}
The components, with which we have to deal with are :  $\delta T_{\phi R\, Ip}  $ and $ \delta T_{\phi \theta\, Ip}$. For the component $ \delta T_{\phi \theta\, Ip} $, the associated components of the four-velocity $u_{\phi} \, , u_{\theta}$ are zero ; and also the associated components of the heat-flux vectors $\gamma_{\theta} \, , \gamma_{\phi}$ are zero. Hence, the $ \delta T_{\phi \theta\, Ip} $ vanishes. On the other hand, for the component $ \delta T_{\phi R \, Ip} $, two among the four terms, which contains $u_{\phi}$ and $\gamma_{\phi}$, vanishes. Hence, 
\begin{equation} \label{6.10}
\delta T_{\phi R\, Ip} =  \gamma_{R} \delta u_{\phi} + u_{R} \delta \gamma_{\phi}  \, . 
\end{equation}   
As a non-zero velocity perturbation in the $\phi $-direction would imply presence of angular momentum in that direction, in the cosmic-fluid being accreted by the black hole, then the accretion would no longer remain spherical and that would result in the formation of accretion-disk around the black hole. So, to avoid this complexity we assume that $ \delta u_{\phi} $ can be neglected. The same argument also holds for the $\gamma_{\phi} $. So, with this assumption $ \delta T_{\phi R} = 0 $. 
Therefore, the equations reduce to :
\begin{equation}  \label{6.14}
\delta_{\omega,q_{2},q_{3}}  R_{ij}  =  0  \, .
\end{equation}  
We begin our calculation from these equations. 
\section{APPENDIX 3 : Determining some essential relations regarding the generalized McVittie metric } \label{RelationsforGMcM} 
In the present work we need various relations between different quantities present in the line-element of the metric i.e. the metric-coefficients and transformation rules to shift from differentiation w.r.t. one coordinate system to the other.
 In this section, we are giving these relations and formula which have been used in our work. 
\\
 The integrating factor $F$ in the equation \ref{1.8} satisfies the differential equation \cite{GaoChen}:
\begin{equation} \label{3.1}
\frac{\partial}{\partial R} \Big( \frac{1}{F} \Big) = \frac{\partial}{\partial t} \Big( \frac{\beta}{F}  \Big) \, , 
\end{equation} 
where $\beta$ is the quantity $\frac{C}{A^{2}-C^{2}}$. To simplify this equation, first of all we express the partial derivative w.r.t. the radial coordinate $R(t,r)$ in Nolan-gauge, in terms of the partial derivative w.r.t. isotropic time coordinate t. As already stated the coordinate $R$ is given by,
\begin{center}
${\displaystyle R = a(t)r \Big( 1 + \frac{M_{H}(t)}{2a(t)r}   \Big)^{2} }  \, .$
\end{center}   
So, the partial derivative of R w.r.t. t is given by :
\begin{center} 
${\displaystyle \frac{\partial R}{\partial t} = \dot{a} r \Big( 1 + \frac{M_{H}(t)}{2a(t)r}  \Big)^{2} + }$
${\displaystyle 2 ar \Big( 1 + \frac{M_{H}(t)}{2a(t)r}  \Big) \frac{1}{2r} \Big( \frac{1}{a}\dot{M}_{H} - \frac{M_{H}}{a^{2}} \dot{a}   \Big)      \, .  }$
\end{center}
(It is quite clear that as the scale-factor a(t) and Hawking-Hayward quasi-local mass $M_{H}(t)$ are the functions of time(t) only, $\partial a/\partial t = da/dt \equiv \dot{a} $ and $\partial M_{H}/\partial t = dM_{H}/dt \equiv \dot{M}_{H} $.) \\On simplifying the above expression of $\partial R / \partial t$, we obtain :
\begin{equation} \label{3.2}
\frac{\partial R}{\partial t } = \left\lbrace HR + M_{H} \Big( 1+\frac{M_{H}}{2ar} \Big) \Big( \frac{\dot{M}_{H}}{M_{H}}- H   \Big) \right\rbrace \, . 
\end{equation}
As, $M_{H}(t) = m(t)a(t)$, it is easy to check that this can be written as :
\begin{equation}  \label{3.3}
\frac{\partial R}{\partial t } =  \left\lbrace HR + \dot{m} a \sqrt{\frac{\tilde{r}}{r}}  \right\rbrace  = C(t,r) \, . 
\end{equation}
Therefore, we can say, 
\begin{equation}  \label{3.4}
\frac{\partial }{\partial R} = \frac{1}{C(t,r)} \frac{\partial}{\partial t} \, . 
\end{equation} 
Hence, the equation \ref{3.1} can be written as :
\begin{equation}  \label{3.5}
\frac{1}{C}\frac{\partial}{\partial t} \Big( \frac{1}{F} \Big) = \frac{\partial}{\partial t} \Big( \frac{\beta}{F}  \Big) \, ,
\end{equation}
\begin{equation} \label{3.6}
{\displaystyle   \Rightarrow  -F \frac{\partial \beta }{\partial F} = \frac{1}{C}- \beta }   \, . 
\end{equation}
Where, \begin{center}
${\displaystyle \frac{1}{C} - \beta  = \frac{1}{C} -  \frac{C}{A^{2}- C^{2}} = \frac{A^{2}- 2 C^{2}}{C(A^{2}- C^{2})}      } \, . $
\vspace{0.5cm}
\end{center}
The above equation \ref{3.6} has to be solved for getting the solution $F$. \\
Again, we have to determine the relation between $ \frac{\partial}{\partial R} $ and $ \frac{\partial}{\partial \overline{t}} $ . 
We have already shown that 
\begin{center}
${\displaystyle   \frac{\partial R}{\partial t} = C(t,R) }\,  ,$
\end{center} 
and the time-coordinate $\overline{t}$, we are working with, is given by : 
\begin{equation} \label{2.1}
d \overline{t} = \frac{1}{F} \Big(  dt + \frac{C}{A^{2}- C^{2}}  dR \Big) \, . 
\end{equation}
From the above equation \ref{2.1} we obtain :
\begin{center}
${\displaystyle \frac{\partial \overline{t}}{\partial R} = \frac{1}{F} \frac{\partial t}{\partial R}  + \frac{C}{F(A^{2}- C^{2}) }   \, ,   }$ 
\end{center}
\begin{equation} \label{2.2}
{\displaystyle  \Rightarrow  \frac{\partial \overline{t}}{\partial R}  =  \frac{1}{FC \Big(  1 - \frac{C^{2}}{A^{2}} \Big)}   \, . }
\end{equation}
Hence, multiplying both sides with the differential operator $\frac{\partial}{\partial \overline{t}} $, we obtain :
\begin{center}
${\displaystyle  \frac{\partial \overline{t}}{\partial R} \frac{\partial}{\partial \overline{t}} =              \frac{\partial }{\partial R} = \frac{1}{FC \Big(  1 - \frac{C^{2}}{A^{2}} \Big)} \frac{\partial}{\partial \overline{t}}  \,  . }$
\end{center}
Inserting the relation between $\frac{\partial}{\partial R} $ and $ \frac{\partial }{\partial t} $, we see :
\begin{equation} \label{2.3}
\frac{1}{C} \frac{\partial }{\partial t} = \frac{1}{FC \Big(  1 - \frac{C^{2}}{A^{2}} \Big) } \frac{\partial }{\partial \overline{t}} \, , 
\end{equation}
or, 
\begin{equation}
 \frac{\partial }{\partial t} = \frac{1}{F \Big(  1 - \frac{C^{2}}{A^{2}} \Big) } \frac{\partial }{\partial \overline{t}} \, . 
\end{equation}
Again, the partial derivative of $R$ w.r.t. $r$ gives :
\begin{equation} \label{2.4}
\begin{aligned}
\frac{\partial R}{\partial r} = a(t) \Big( 1 + \frac{M_{H}(t)}{2a(t)r}   \Big)^{2} +
 2a(t)r \Big( 1 + \frac{M_{H}(t)}{2a(t)r}   \Big) \Big(  -\frac{M_{H}(t)}{2a(t)r^{2}} \Big)   
\\
= a(t) \Big( 1 + \frac{M_{H}(t)}{2a(t)r} \Big) \Big( 1 - \frac{M_{H}(t)}{2a(t)r} \Big) 
\\
= a(t) \Big( 1 - \frac{M_{H}^{2}(t)}{(2a(t)r)^{2}} \Big) \, . 
\end{aligned}
\end{equation}
So, the relation between the partial derivatives w.r.t. $R$ and $r$ can be written as : 
\begin{equation} \label{2.5}
\frac{\partial }{\partial r} = a(t) \Big( 1 - \frac{M_{H}^{2}(t)}{(2a(t)r)^{2}} \Big) \frac{\partial }{\partial R} \, . 
\end{equation}
\section{APPENDIX 4 : Some relations regarding Fourier-transformation and Convolution theorem, which have been applied in the work } \label{Fourier-transform}
It is to be noted that every term appearing in the integrands in the equations \ref{4.20a} and \ref{4.21a} can be written as a multiplication of an unperturbed term (containing unperturbed quantities appearing in the metric-coefficients) and a perturbation term. While some of the perturbation terms have partial derivatives w.r.t. $t$ acting on the perturbations. Considering these two type of terms separately, we first write their Fourier-integral versions, inside the overall Fourier-integral from time-space to frequency-space. Thereafter, the partial-derivatives of $t$ acting on the perturbations will give rise to a factor of $i\sigma ''$, where $\sigma'' $ is the integrating variable and consequently $(i\sigma '')^{2}$. We shall consider the perturbation term with those factors as a whole. Thereafter, we apply the `Covolution theorem' for Fourier-transformations for the product of these two types of terms viz. the unperturbed term and the perturbation term. According to the `Covolution theorem', one of these terms, say the perturbation term will be a function pf $(\sigma - \sigma ')$ inside the integral, where $\sigma$ is the integrating variable and the Fourier-transform is evaluated at the frequency $\sigma'$. We take the case of $\sigma' = 0 $, which actually makes the Fourier-integral proportional to the mean of the integrand. Then, we equate the integrands from both sides of the equations.\\
We describe the procedure in a generalized way. Suppose, we have the equation : 
\begin{equation} \label{a1}
f(r, \theta, t) \delta \alpha(r, \theta, t) = g(r, \theta, t) \delta \beta(r, \theta, t) \, , 
\end{equation}
where $f(r, \theta, t)$ and $g(r, \theta, t)$ consists of unperturbed parameters, present in the metric coefficients and functions of $r \, , \theta $ and $t$. $ \delta \alpha $ and $ \delta \beta $ are linear perturbations, which are all time($ t $)-varying . The aim is to map this equation from time-space to frequency-space employing Fourier-transformation. \\
To map the time-dependence of the overall equation from time-space to frequency-space we fourier-transform both sides of the equation directly. This gives :
\begin{equation} \label{a2}
\int_{0}^{\infty}  f(t) \delta \alpha(t) \, e^{-i \sigma' t}  dt =  \int_{0}^{\infty}  g(t) \delta \beta(t) \, e^{-i \sigma' t}  dt \, .
\end{equation}
Now applying the `Convolution theorem' for Fourier-transforms, we write the equation \ref{2.1} as :
\begin{equation} \label{a3}
\int_{-\infty}^{\infty}  f^{\dagger}(\sigma - \sigma') \delta \alpha^{\dagger} (\sigma) d\sigma =  \int_{-\infty}^{\infty}  g^{\dagger}(\sigma - \sigma') \delta \beta^{\dagger} (\sigma) d\sigma \, . 
\end{equation}
If we take the case of $\sigma' = 0 $ for the equation \ref{2.1}, then it gives : 
\begin{equation} \label{a4}
\int_{-\infty}^{\infty} f^{\dagger}(\sigma) \delta \alpha^{\dagger} (\sigma) d\sigma =  \int_{-\infty}^{\infty} g^{\dagger}(\sigma ) \delta \beta^{\dagger} (\sigma) d\sigma \, .  
\end{equation}
Considering the case of $\sigma' = 0 $ physically means that the Fourier-integral becomes proportional to the mean of the integrand. Thereafter equating the integrands of the integrals from both sides of the equation \ref{a4} (as this can be done without any loss of generality), we get :
\begin{equation} \label{a5}
 f^{\dagger}(\sigma) \delta \alpha^{\dagger} (\sigma) =  g^{\dagger}(\sigma) \delta \beta^{\dagger} (\sigma) \, . 
\end{equation}
Again, we consider that the LHS of the equation \ref{a1} can also be written as $\tilde{f}\delta \tilde{\alpha} $ i.e. $ \tilde{f}\delta \tilde{\alpha} = f \delta \alpha $. This is nothing but we define the linear perturbation in a different way. Then, the Fourier-transform of both $ \tilde{f}\delta \tilde{\alpha} $ and $ f \delta \alpha $ will be same viz. :
\begin{equation} \label{a7}
\int_{0}^{\infty}  f(t) \delta \alpha(t) \, e^{-i \sigma' t}  dt = \int_{0}^{\infty}  \tilde{f}(t) \delta \tilde{\alpha}(t) \, e^{-i \sigma' t}  dt \, . 
\end{equation}
Applying the Convolution theorem, similarly as we did for getting equation \ref{a3} from \ref{a2}, and then taking the case of zero-frequency i.e. $\sigma' = 0 $, from the equation \ref{a7} we get : 
\begin{equation} \label{a8}
\int_{-\infty}^{\infty}  f^{\dagger}(\sigma ) \delta \alpha^{\dagger}(\sigma) d\sigma = \int_{-\infty}^{\infty} \tilde{f}^{\dagger}(\sigma) \delta \tilde{\alpha}^{\dagger}(\sigma) d\sigma \, ,  
\end{equation}
which in turn gives :
\begin{equation} \label{a9}
f^{\dagger}(\sigma ) \delta \alpha^{\dagger}(\sigma) = \tilde{f}(\sigma)^{\dagger} \delta \tilde{\alpha}^{\dagger}(\sigma) \, ,
\end{equation}
by equating the integrands. The result of equation \ref{a9} is not only valid for the case of a term with perturbation, but also for any function i.e. say if $f = f_{1}(t)f_{2}(t) = \tilde{f}_{1}(t) \tilde{f}_{2}(t) $, then it implies $f_{1}^{\dagger}(t)f_{2}^{\dagger}(t) = \tilde{f}_{1}^{\dagger}(t) \tilde{f}_{2}^{\dagger}(t) $. \\
Now, we analyse the case where the functions $f(t)$ and $g(t)$ contains partial differential-operators w.r.t. $t$ in general, as this is actually the case of perturbation equations in our work. We write $\hat{f}(t)$ as $ \hat{f}(t) \equiv f(t) + h(t) \frac{\partial}{ \partial t} $. So, the Fourier-transform of the part $f(t)\delta \alpha(t) $ proceeds in the usual way, as has been shown. The  quantity $ h(t) \frac{\partial}{ \partial t} (\delta \alpha) $ is of our interest here. Its Fourier-transform gives : 
\begin{equation} \label{a10}
\begin{aligned}
\left\lbrace \left( h(t) \frac{\partial}{ \partial t}  \right)  \delta \alpha \right\rbrace^{\dagger}  = \frac{1}{2\pi} \int_{0}^{\infty} \left(  h(t) \frac{\partial}{ \partial t}  \right)  \delta \alpha (t)  \, e^{-i \sigma' t}  dt  \, .
\end{aligned}
\end{equation}
\begin{widetext}
Substituting $h(t)$ and $\delta \alpha(t) $ in terms of their Fourier-transforms from time-space to frequency-space in the integrand on the RHS of the above equation \ref{a10}, we get : 
\begin{equation} \label{a11}
\begin{aligned}
\left\lbrace \left( h(t) \frac{\partial}{ \partial t} \right)  \delta \alpha \right\rbrace ^{\dagger} = \left( \frac{1}{2\pi}\right)^{3}  \int_{0}^{\infty} \left(  \int_{-\infty}^{\infty} h^{\dagger}(\sigma) e^{i \sigma t}  d\sigma \right)  \frac{\partial}{ \partial t} \left( \int_{-\infty}^{\infty} \delta \alpha^{\dagger} (\sigma'') e^{i \sigma'' t}  d\sigma'' \right)  \, e^{-i \sigma' t}  dt  \, , 
\\
=  \left( \frac{1}{2\pi}\right)^{3}  \int_{0}^{\infty} \left(  \int_{-\infty}^{\infty} h^{\dagger}(\sigma) e^{i \sigma t}  d\sigma \right)  \left( \int_{-\infty}^{\infty} i(\sigma'' - \sigma') \delta \alpha^{\dagger} (\sigma'') e^{i \sigma'' t}  d\sigma'' \right)  \, e^{-i \sigma' t}  dt  \, , 
\\
\Rightarrow \left\lbrace \left( h(t) \frac{\partial}{ \partial t} \right)  \delta \alpha \right\rbrace ^{\dagger} = \left( \frac{1}{2\pi}\right)^{3}  \int_{0}^{\infty}  \int_{-\infty}^{\infty} h^{\dagger}(\sigma)   \int_{-\infty}^{\infty} i(\sigma'' - \sigma') \delta \alpha^{\dagger} (\sigma'')   \, e^{i (\sigma +\sigma''-\sigma' ) t} dt d\sigma   d\sigma''   \,  . 
\end{aligned}
\end{equation}
\end{widetext}
The RHS of the above equation \ref{a11} contains the integral of $ e^{i (\sigma +\sigma''-\sigma' ) t } $ over time. This is just the Fourier-transform of a plane-wave, which is actually the Dirac-Delta function viz. :
\begin{equation} \label{a12}
 \frac{1}{2\pi} \int_{0}^{\infty}   e^{i (\sigma +\sigma''-\sigma' ) t}  dt = \delta(\sigma +\sigma''-\sigma')
 =  \delta(\sigma'-\sigma - \sigma'') \, . 
\end{equation}  
\footnote{The last step uses the property that Dirac-Delta function is symmetric. }
So, we write the equation \ref{a11} as :
\begin{equation}  \label{a13}
\begin{aligned}
\left\lbrace \left( h(t) \frac{\partial}{ \partial t} \right)  \delta \alpha \right\rbrace^{\dagger}=
 \left( \frac{1}{2\pi}\right)^{2}  \int_{-\infty}^{\infty} \int_{-\infty}^{\infty}   h^{\dagger}(\sigma) \,  i(\sigma'' - \sigma') \delta \alpha^{\dagger} (\sigma'') 
 \\
   \delta(\sigma'-\sigma - \sigma'') d\sigma   d\sigma''   \,  .
\end{aligned}
\end{equation}
Using the property of Dirac-Delta function on the RHS of the equation \ref{a13}, integrating over $\sigma'' $, we get : 
\begin{equation} \label{a14}
\begin{aligned}
\left\lbrace \left( h(t) \frac{\partial}{ \partial t} \right)  \delta \alpha \right\rbrace^{\dagger}= 
\\
\left( \frac{1}{2\pi}\right)^{2}  \int_{-\infty}^{\infty}   h^{\dagger}(\sigma) \,  i(\sigma'-\sigma - \sigma') \delta \alpha^{\dagger} (\sigma'')  d\sigma    \,
\\
 =  \left( \frac{1}{2\pi}\right)^{2}  \int_{-\infty}^{\infty}   h^{\dagger}(\sigma) \, (-i\sigma) \delta \alpha^{\dagger} (\sigma'- \sigma)  d\sigma    \,  .
\end{aligned}
\end{equation}
Now, if we replace $\sigma $ with $- \sigma $ in the integral on RHS of the above equation, then we shall obtain : 
\begin{equation} \label{a15}
\left\lbrace \left( h(t) \frac{\partial}{ \partial t} \right)  \delta \alpha \right\rbrace^{\dagger} =
\left( \frac{1}{2\pi}\right)^{2}  \int_{-\infty}^{\infty}   h^{\dagger}(-\sigma) \, (+i\sigma) \delta \alpha^{\dagger} (\sigma'+ \sigma)  d\sigma  \, . 
\end{equation}
Then for the case of $ \sigma' = 0 $, this gives : 
\begin{equation} 
\left\lbrace \left( h(t) \frac{\partial}{ \partial t} \right)  \delta \alpha \right\rbrace^{\dagger} =
\left( \frac{1}{2\pi}\right)^{2}  \int_{-\infty}^{\infty}   h^{\dagger}(-\sigma) \, (+i\sigma) \delta \alpha^{\dagger} (\sigma)  d\sigma  \, . 
\end{equation}
\section{APPENDIX 5 : Analyzing the quantities $\protect \frac{(\mathbb{F}(r,t)+ i \sigma )}{(\mathscr{F}(r,t)+ i \sigma )}$, $\protect \frac{\mathbb{F}(r,t)}{\mathscr{F}(r,t) } $ and the approximations applicable to these }  
\label{Appendix-4}
\subsection{Expressing the ratio $\protect \frac{\mathbb{F}(r,t)}{\mathscr{F}(r,t) } $ conveniently in terms of $\protect C, \, A, \Delta \, , F $ and their derivatives :}
In this sub-section, we are going to express the quantities $ \mathbb{F}$ and $ \mathscr{F}$ in terms of $ C, \, A, \Delta \, , F $ and their derivatives ; and then in the next sub-section we shall investigate some approximations, which will be applicable to our calculations.  \\
The quantity $ \frac{\mathbb{F}(r,t)}{\mathscr{F}(r,t) }$ can be expressed as :
\begin{equation} \label{7.1}
\frac{\mathbb{F}(r,t)}{\mathscr{F}(r,t) }  = \frac{\frac{\Delta}{R^{4}} \Big( 1 - \frac{C^{2}}{A^{2}}  \Big) F  \frac{\partial}{\partial \overline{t}} \left\lbrace  \frac{\Delta}{R^{4}} \Big( 1 - \frac{C^{2}}{A^{2}}  \Big) F  \right\rbrace^{-1} }{\frac{F}{R^{4}} \frac{\partial}{\partial \overline{t}} \Big( \frac{R^{4}}{F}  \Big)}  \, , 
\end{equation}
\begin{equation} \label{7.2}
\begin{aligned}
\Rightarrow   \frac{\mathbb{F}(r,t)}{\mathscr{F}(r,t) } 
= 1 + \frac{\Delta \Big( 1 - \frac{C^{2}}{A^{2}} \Big) \frac{\partial}{\partial \overline{t}} \Big( \Delta \Big( 1 - \frac{C^{2}}{A^{2}} \Big)  \Big)^{-1} }{ \frac{F}{R^{4}} \frac{\partial}{\partial \overline{t}} \Big( \frac{R^{4}}{F}  \Big)} \, . 
\end{aligned} 
\end{equation}
The denominator in the additional term with 1 on the RHS of the above equation \ref{7.2} can be written as :
\begin{equation} \label{7.3}
\begin{aligned}
 \frac{1}{\frac{F}{R^{4}} \frac{\partial}{\partial \overline{t}} \Big( \frac{R^{4}}{F}  \Big)} 
= \left\lbrace   F \frac{\partial}{\partial \overline{t}} \Big( \frac{1}{F} \Big) +    \frac{1}{R^{4}} \frac{\partial}{\partial \overline{t}} R^{4} \right\rbrace^{-1}   \, . 
\end{aligned}
\end{equation} 
Using the differential relation satisfied by F given in \ref{3.1} we can obtain the following relation :
\begin{equation} \label{7.4}
F \left\lbrace  \frac{\partial}{\partial \overline{t}} \Big( \frac{1}{F}  \Big) \right\rbrace =  \Big( \frac{1}{C}- \beta \Big)^{-1}  \frac{\partial \beta }{\partial \overline{t}} \, , 
\end{equation}
where the quantity $\beta $ is given by : $\beta = \frac{C}{A^{2} - C^{2}} $ ;
and 
using the relations between $\frac{\partial}{\partial R}$ and $\frac{\partial}{\partial \overline{t}}  $, derived in the appendix 3 i.e. section \ref{RelationsforGMcM}, we can easily get :
\begin{equation} \label{7.5}
\frac{1}{R^{4}} \frac{\partial}{\partial \overline{t}} R^{4} = \frac{4}{R} FC \Big( 1 - \frac{C^{2}}{A^{2}} \Big) \, . 
\end{equation} 
We write the detailed expressions of $\frac{\partial \beta}{\partial \overline{t}} $ and $ \frac{\partial A}{\partial \overline{t}} $ respectively as :
\begin{equation} \label{7.6}
\frac{\partial \beta}{\partial \overline{t}}  = \frac{1}{(A^{2} - C^{2})^{2}} \left\lbrace (A^{2} + C^{2}) \frac{\partial C}{\partial \overline{t}}  - 2 CA \frac{\partial A}{\partial \overline{t}}\right\rbrace \, . 
\end{equation}
and
\begin{equation} \label{7.7}
\begin{aligned}
{\displaystyle \frac{\partial A}{\partial \overline{t}} = \frac{\partial}{\partial \overline{t} } \Big(  1 -  \frac{2M_{H}}{R} \Big)   
  = -2 \left\lbrace  - \frac{M_{H} }{R^{2}} \frac{\partial R}{\partial \overline{t}}  + \frac{1}{R} \frac{\partial M_{H}}{\partial \overline{t}} \right\rbrace   }\, . 
  \end{aligned}
\end{equation}
Here, $\frac{\partial M_{H}}{\partial \overline{t}} $ can be expressed as :
\begin{center} 
${\displaystyle  
\frac{\partial M_{H}}{\partial \overline{t}}  = F \Big(  1 - \frac{C^{2}}{A^{2}} \Big)  \frac{\partial M_{H}}{\partial t }    }$
${\displaystyle = M_{H}F \Big(  1 - \frac{C^{2}}{A^{2}} \Big) \Big( H + \frac{\dot{m}}{m} \Big) \, .   }$
\end{center}
Substituting the expression of the $\frac{\partial R}{\partial \overline{t}}$ in the equation \ref{7.7}, we obtain :
\begin{equation} \label{7.8a}
{\displaystyle \frac{\partial A}{\partial \overline{t}} = -2 \frac{M_{H} F}{R^{2}} \Big(  1 - \frac{C^{2}}{A^{2}} \Big) \left\lbrace  -C + HR + R \frac{\dot{m}}{m} \right\rbrace   } \, . 
\end{equation}  
As, $C = HR +  \dot{m} a \sqrt{\frac{\tilde{r}}{r}}$, substituting it in the RHS of the above equation \ref{7.8a} we obatin :
\begin{equation}  \label{7.8b}
\begin{aligned}
{\displaystyle  \frac{\partial A}{\partial \overline{t}}  =  -2 \frac{M_{H} F}{R^{2}} \dot{m} a \Big(  1 - \frac{C^{2}}{A^{2}} \Big) \left\lbrace   - \sqrt{\frac{\tilde{r}}{r}} + \frac{\tilde{r}}{m}   \right\rbrace         } \, . 
\end{aligned}
\end{equation}
Substituting the expressions :
\begin{center}
${\displaystyle    \sqrt{\frac{\tilde{r}}{r} } = \Big( 1 + \frac{M_{H}}{2  r a }  \Big) = \Big( 1 + \frac{m}{2  r  }  \Big)  }$ 
\vspace{0.2cm}\\
and  \vspace{0.2cm}\\
${\displaystyle \frac{\tilde{r}}{m}  = \frac{r}{m} \Big( 1 + \frac{m}{2  r  }  \Big)^{2}  }$
${\displaystyle =  \Big( \frac{r}{m} +1+ \frac{m}{4r}  \Big) } \, , $ 
\end{center}
on the RHS of the equation \ref{7.8b}, we get :
\begin{equation}  \label{7.9}
\frac{\partial A}{\partial \overline{t}} = -2 \frac{M_{H} F}{R^{2}} \dot{m} a \Big(  1 - \frac{C^{2}}{A^{2}} \Big)\left(  \frac{r}{m}  - \frac{m}{4r} \right) \, . 
\end{equation}
Calculating the detailed expression of the quantity $\frac{\partial C}{\partial \overline{t}} $ we obtain :
\begin{equation} \label{7.10}
\begin{aligned}
\frac{\partial C}{\partial \overline{t}} = F \Big(  1 - \frac{C^{2}}{A^{2}} \Big) 
\\
\left\lbrace  - H^{2} R + 2 \dot{m} \dot{a} \Big( 1+ \frac{m}{2r} \Big) + \frac{\dot{m}^{2} a}{ 2 r} + \ddot{m} a \Big( 1+ \frac{m}{2r} \Big) \right\rbrace  \, . 
\end{aligned}
\end{equation}
Hence, the denominator in the additive term with 1 on the RHS of the equation \ref{7.2} can be written as :
\begin{center}
${\displaystyle
 \frac{F}{R^{4}}\frac{\partial }{\partial \overline{t}} \Big(  \frac{R^{4}}{F}  \Big) = 
\Big(  \frac{1}{C} - \beta \Big)^{-1} \frac{\partial \beta }{\partial \overline{t}} + \frac{4}{R} FC \Big(  1 - \frac{C^{2}}{A^{2}} \Big)  
}$  
${\displaystyle = C \Big(  1 - \frac{C^{2}}{A^{2}} \Big)  \left\lbrace \Big( 1 -  \frac{2C^{2}}{A^{2}} \Big)^{-1}  \frac{\partial \beta }{\partial \overline{t}} +  \frac{4}{R} F \right\rbrace   }\, . $
\end{center}
While the numerator of that term is given by : 
\begin{center}
${\displaystyle  \Delta  \Big(  1 - \frac{C^{2}}{A^{2}} \Big)  \frac{\partial}{\partial \overline{t}} \Big(\Delta  \Big(  1 - \frac{C^{2}}{A^{2}} \Big) \Big)^{-1} 
}$ \, . 
\end{center}
Hence, the additive term with 1 on the RHS of the equation \ref{7.2} can be written as : 
\begin{equation}
\begin{aligned}
 \frac{\Delta \Big( 1 - \frac{C^{2}}{A^{2}} \Big) \frac{\partial}{\partial \overline{t}} \Big( \Delta \Big( 1 - \frac{C^{2}}{A^{2}} \Big)  \Big)^{-1} }{ \frac{F}{R^{4}} \frac{\partial}{\partial \overline{t}} \Big( \frac{R^{4}}{F}  \Big)} 
 \\
 = -\frac{\Big( 1 - \frac{C^{2}}{A^{2}} \Big)^{-2}}{\Delta C } \frac{\frac{\partial}{\partial \overline{t}} \Big(\Delta  \Big(  1 - \frac{C^{2}}{A^{2}} \Big) \Big)^{-1} }{\left\lbrace \Big( 1 -  \frac{2C^{2}}{A^{2}} \Big)^{-1}  \frac{\partial \beta }{\partial \overline{t}} +  \frac{4}{R} F \right\rbrace } \, . 
\end{aligned}  
\end{equation}
\begin{widetext}
Writing the detailed expressions of numerator and denominator in the additive term with 1 on the RHS of the equation \ref{7.2}, in terms of $ C, \, A, \, \Delta, \, F $ and their derivatives w.r.t. $ \overline{t} $, we obtain :
\begin{equation}  \label{7.12}
\begin{aligned}
\frac{{\displaystyle \Delta \Big( 1 - \frac{C^{2}}{A^{2}} \Big) \frac{\partial}{\partial \overline{t}} \Big( \Delta \Big( 1 - \frac{C^{2}}{A^{2}} \Big)  \Big)^{-1} }}{ {\displaystyle \frac{F}{R^{4}} \frac{\partial}{\partial \overline{t}} \Big( \frac{R^{4}}{F}  \Big)  }}  
=  
 \frac{  {\displaystyle  \left\lbrace \Big( \frac{-2}{A^{2}} \frac{\partial C }{\partial \overline{t}} + \frac{2 C}{A^{3}} \frac{\partial A }{\partial \overline{t}}   \Big)  + \Big(  1 - \frac{C^{2}}{A^{2}} \Big) \frac{1}{\Delta} \frac{\partial \Delta }{\partial \overline{t}} \right\rbrace } }{ {\displaystyle  \Big( 1 -  \frac{2C^{2}}{A^{2}} \Big)^{-1}  \left\lbrace \Big( 1 + \frac{C^{2}}{A^{2}} \Big) \frac{1}{A^{2}}  \frac{\partial C }{\partial \overline{t}} - \frac{2C}{A^{3}} \frac{\partial A }{\partial \overline{t}}  \right\rbrace   +  \Big( 1 - \frac{C^{2}}{A^{2}} \Big)^{2} \frac{4}{R} F  } }    \, .
\end{aligned}
\end{equation} 
\end{widetext}
\subsection{Checking the order of different quantities present in the ratio $ \frac{\mathbb{F}(r,t)}{\mathscr{F}(r,t) }  $ and applying approximation :  } \label{Approximations}
If we examine the ratio given in equation \ref{7.12} in the previous sub-section, then in the numerator and denominator of the ratio on the RHS of the equation \ref{7.12}, the quantities $ \Big( \frac{-2}{A^{2}} \frac{\partial C }{\partial \overline{t}} + \frac{2 C}{A^{3}} \frac{\partial A }{\partial \overline{t}}   \Big)  $ and $ \Big( 1 -  \frac{2C^{2}}{A^{2}} \Big)^{-1}  \left\lbrace \Big( 1 + \frac{C^{2}}{A^{2}} \Big) \frac{1}{A^{2}}  \frac{\partial C }{\partial \overline{t}} - \frac{2C}{A^{3}} \frac{\partial A }{\partial \overline{t}}  \right\rbrace $ should have same order of magnitude within finite distance from the black hole \footnote{It is quite clear that the quantities $ \Big( 1 -  \frac{2C^{2}}{A^{2}} \Big) $, $ \Big( 1 + \frac{C^{2}}{A^{2}} \Big) $ and $ \Big( 1 - \frac{C^{2}}{A^{2}} \Big) \, \sim 1 $ within finite distance from the primordial black hole described by the generalized McVittie metric.}. Therefore, the order of the magnitude of the ratio given in equation \ref{7.12} would depend mainly on the quantities $\frac{1}{\Delta} \frac{\partial \Delta }{\partial \overline{t}} $ and $ F/R $, if their order of magnitude is higher than the former quantities. \\
Therefore, now we have to check the relative significance of different quantities in the ratio given in equation \ref{7.12} for having an idea on its overall order of magnitude. \\
First of all, we check the relative significance of the quantities : $\Big( 1 - \frac{C^{2}}{A^{2}} \Big) \frac{1}{\Delta} \frac{\partial \Delta }{\partial \overline{t}} $ and $ \Big( 1 - \frac{C^{2}}{A^{2}} \Big)^{2} \frac{4F}{R} $. Their ratio can be expressed as :
\begin{equation} \label{7.14}
\begin{aligned}
\frac{   {\displaystyle \Big( 1 - \frac{C^{2}}{A^{2}} \Big) \frac{1}{\Delta} \frac{\partial \Delta   }{\partial \overline{t}} }  }{  {\displaystyle \Big( 1 - \frac{C^{2}}{A^{2}} \Big)^{2} \frac{4F}{R}  }}  = 
 \frac{   {\displaystyle   \frac{2C(R- M_{H})- 2 \dot{M}_{H}R }{R^{2} - 2 M_{H} R }}}{ {\displaystyle \frac{4}{R} }  } 
\\
= \frac{1}{2} \left\lbrace  C\frac{ \Big(  1- \frac{M_{H}}{R} \Big)  }{ \Big(  1 - 2 \frac{M_{H}}{R}  \Big)}   - \frac{ \dot{M}_{H}}{ \Big( 1 -  2 \frac{M_{H}}{R}   \Big)  }  \right\rbrace \, . 
\end{aligned}
\end{equation} 
Hence, if magnitude of $C << 1 $ and magnitude of $\dot{M}_{H} << 1$, then the magnitude of the above ratio is also $<< 1$. \\
To establish the fact that the order of magnitude of $C$ and $\dot{M}_{H}$ are $<< 1 $, we plot these w.r.t. time from $10^{-25} \, s$ to $100 \, s$ after Big-bang. We separately plot the two parts of $C $ viz. $HR/c$ and $ \frac{G}{c^{3}} \dot{m} a \sqrt{\frac{\tilde{r}}{r}} $ ; as these vary with time in different ways. 
 It is to be noted that the radial-distance $R$ is arbitrary in this case, but that does not imply that it can be taken to theoretically-infinite or asymptotic distance. This is because at theoretically-infinite distance the generalized McVittie metric would reduce to FLRW-metric. The evolution of perturbations, around the PBH, is mainly to be studied within finite distance from the PBH. To set a characteristic length-scale for plotting $C $ w.r.t. time, we use the comoving Schwarzschild length-scale $R_{s} = \frac{2 G M_{H}}{c^{2}} $ (not the horizon), for the PBHs of maximum mass available at a certain instant of time in early Universe, which is just the horizon-mass $m_{h}$ at that time. The horizon-mass $m_{h}$ at a time t seconds after Big-bang is given by $ m_{h} = 10^{35}t \, kg $ \cite{, Carr1, Carr2}. In this way, we depict the maximum possible value of the comoving Schwarzschild length-scale $R_{s}$ at that time. The plot of $HR_{s}/c $ w.r.t. time has been shown in the figure \ref{C1vstimePlot}.  
\begin{figure}[h]
\includegraphics[width=7.4cm]{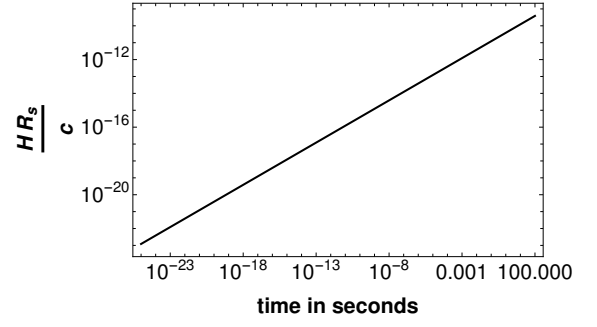}
\caption{Plot in log-log scale, showing the variation of $\frac{HR_{s}}{c}$ with time from $10^{-25}\, s $ to $100\, s $ after Big-bang. 
 }\label{C1vstimePlot}
\end{figure}
\begin{figure}[h]
\includegraphics[width=7.4cm]{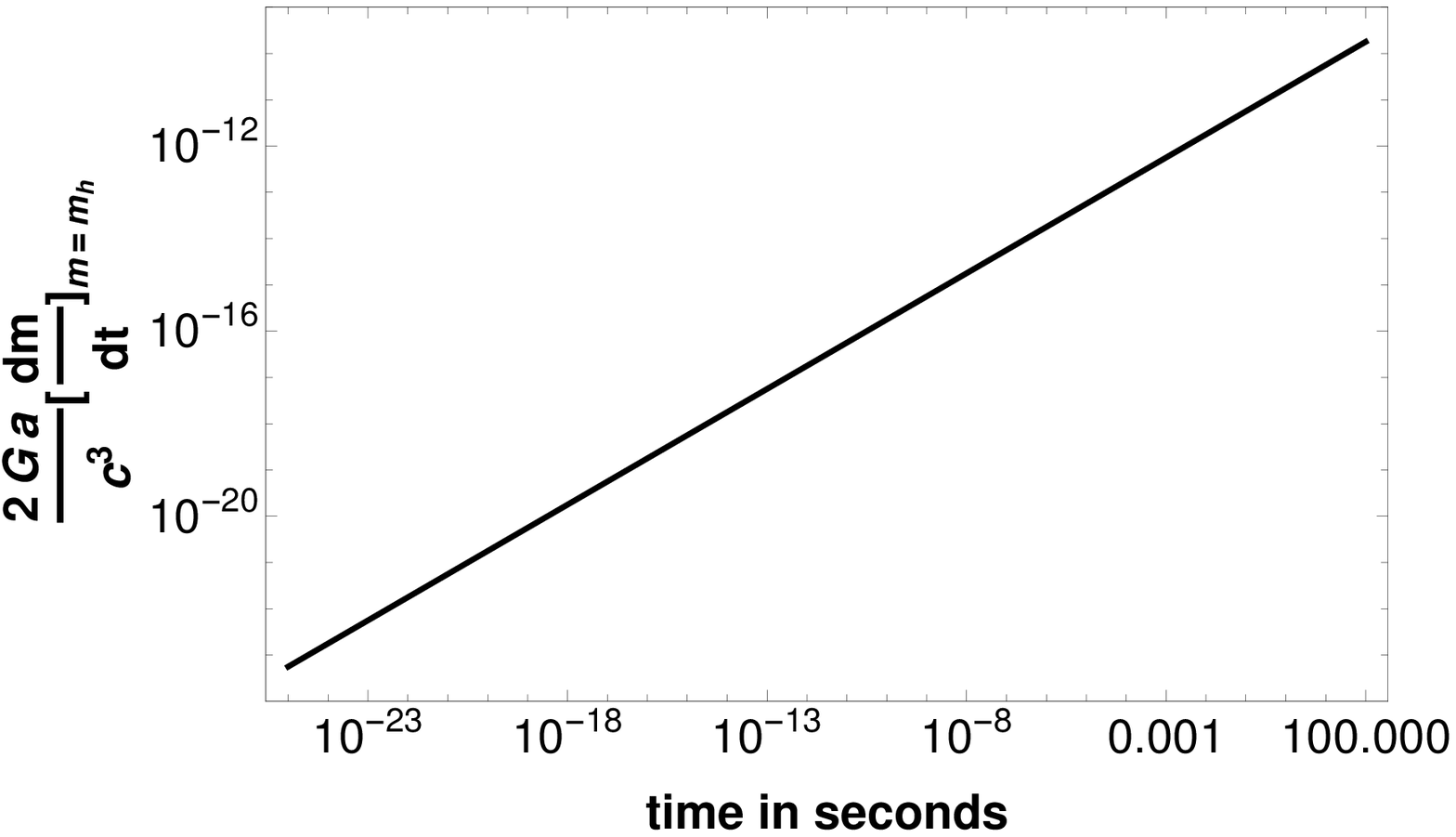}
\caption{Plot in log-log scale, showing the variation of $\frac{2 G}{c^{3}} a [\dot{m}]_{m=m_{h}}$ with time from $10^{-25}\, s $ to $100\, s $ after Big-bang.
 } \label{C2vstimePlot}
\end{figure}

On the other-hand, for $ \frac{G}{c^{3}} \dot{m} a \sqrt{\frac{\tilde{r}}{r}} $, the quantity $ \sqrt{\frac{\tilde{r}}{r}} $ for the Schwarzschild length-scale would be 2 ; as for the Schwarzschild radius, isotropic radial-coordinate $r = \frac{Gm}{2c^{2}} $. Again, we use $\dot{m} $ at $m = m_{h} $, thereby taking the maximum possible value of PBH-mass at a certain instant of time. The plot of $\frac{2 G}{c^{3}} a [\dot{m}]_{m=m_{h}}$ w.r.t. time has been shown in the figure \ref{C2vstimePlot}. In this case, it is worth mentioning that here we are considering the mass-range of PBHs such that their mass-change due to Hawking-evaporation would be negligible and the only significant way of mass-change is the spherical accretion of the surrounding radiation.  \\
Another issue is to be noted in these cases, that we have shown these plots for time till 100 s after Big-bang, mainly because this is the order of time (at which Big-bang nucleosynthesis occurred), at which new PBH-production, by direct gravitational-collapse of sufficiently deep density-perturbations, is predicted to be stopped. 
\vspace{0.2cm}\\
Next, we show the plot of $\frac{G}{c^{3}}[\dot{M_{H}}]_{m=m_{h}} $, which appears in the second term of the expression, on the RHS of the equation \ref{7.14}, w.r.t. time in the figure \ref{MHdotvstime}. \footnote{One part in $\frac{G}{c^{3}}[\dot{M_{H}}]_{m=m_{h}} $ viz. the $\frac{G}{c^{3}}a[\dot{m}]_{m=m_{h}} $, has already been plotted in figure \ref{C2vstimePlot}, with the factor 2. So, if the other part $ \frac{G}{c^{3}}\dot{a}m_{h} $ is less than or equal to the former, then the order of the whole $\frac{G}{c^{3}}[\dot{M_{H}}]_{m=m_{h}} $ would be same as that of $\frac{G}{c^{3}}a[\dot{m}]_{m=m_{h}} $ ; and in fact this is happening in this case. } \\
\begin{figure}
\includegraphics[width=8.3cm]{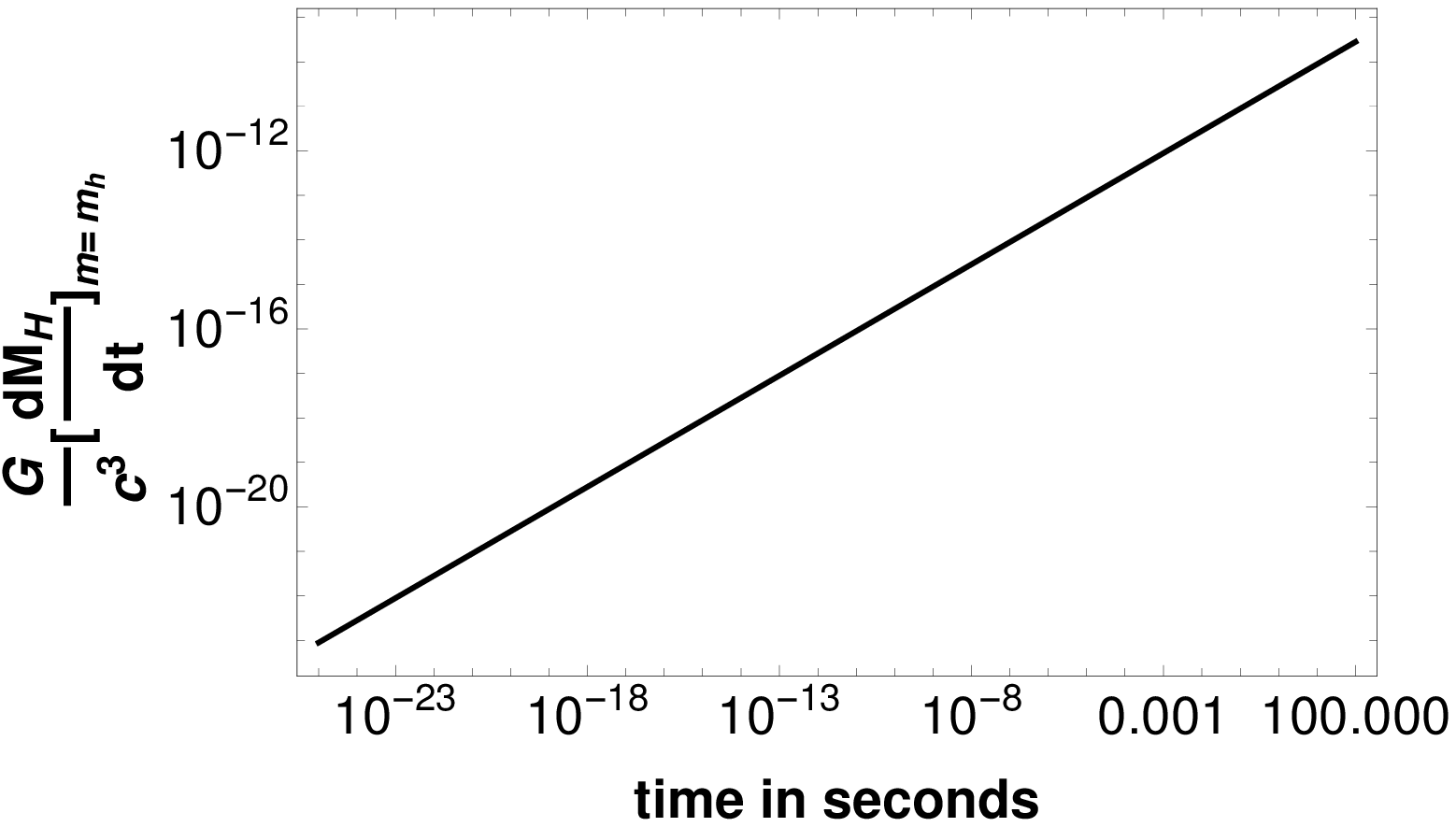}
\caption{Plot in log-log scale, showing the variation of $\frac{G}{c^{3}}[\dot{M_{H}}]_{m=m_{h}} $ with time from $10^{-25}\, s $ to $100\, s $ after Big-bang.
 } \label{MHdotvstime}
\end{figure}
So, we see that all the three quantities $\frac{HR_{s}}{c}$, $\frac{2 G}{c^{3}} a [\dot{m}]_{m=m_{h}}$ and $\frac{G}{c^{3}}[\dot{M_{H}}]_{m=m_{h}} $ have magnitudes within order of $10^{-23} $ to $10^{-9} $, for the range of time from $10^{-25}\, s $ to $100\, s $ after Big-bang. 
\footnote{Another issue is also to be noted that the overall order of the ratio given in equation \ref{7.14} may be lesser (even before $ 10^{-23} \, s $) as in the expression derived in equation \ref{7.14}, the term containing $\dot{M}_{H} $ is subtracted from the term containing $ C $.} 
\\
Therefore, from these plots, it is clear that the approximation based on $ C << 1 $ and $\dot{M}_{H} << 1 $ are very well valid in the specified range of time.     
We now check the relative significance of the quantities :  $\Big( 1 - \frac{C^{2}}{A^{2}} \Big) \frac{1}{\Delta} \frac{\partial \Delta }{\partial \overline{t}} $ and $ \Big( \frac{-2}{A^{2}} \frac{\partial C }{\partial \overline{t}} + \frac{2 C}{A^{3}} \frac{\partial A }{\partial \overline{t}}   \Big) $.
\begin{widetext}
 We take their ratio : 
\begin{equation} \label{7.15}
\begin{aligned}
\frac{ \Big( \frac{-2}{A^{2}} \frac{\partial C }{\partial \overline{t}} + \frac{2 C}{A^{3}} \frac{\partial A }{\partial \overline{t}}   \Big) }{ \Big( 1 - \frac{C^{2}}{A^{2}} \Big) \frac{1}{\Delta} \frac{\partial \Delta }{\partial \overline{t}}  }  
= 
 \frac{2}{A^{2}} \left[ \frac{- \left\lbrace  H^{2}R + \big( 1 + \frac{m}{2r}  \big)(2 \dot{m} \dot{a} + \ddot{m} a )  + \frac{\dot{m}^{2} a }{ 2 r }    \right\rbrace  +  \frac{C}{A} \left\lbrace - 2 \frac{M_{H}}{R^{2}} \dot{m} a  \Big( 1 + \frac{m}{2r} \Big) \Big( \frac{r}{m} - \frac{1}{2} \Big) \right\rbrace  }{\frac{1}{R} \left\lbrace  C\frac{ \Big(  1- \frac{M_{H}}{R} \Big)  }{ \Big(  1 - 2 \frac{M_{H}}{R}  \Big)}   - \frac{ \dot{M}_{H}}{ \Big( 1 -  2 \frac{M_{H}}{R}   \Big)  }  \right\rbrace   } \right]  \,  . 
\end{aligned}   
\end{equation}
\end{widetext}
Now, it is to be noted that in the ratio given in the above equation \ref{7.15}, the quantities containing $\dot{m}, \, \ddot{m} $ or $\dot{M}_{H} $, have the constant $ Gc^{-3}$ with each of them, as we have described in the APPENDIX-1 viz. section \ref{DimensionalIssue}. 
Furthermore, these quantities have $ a $ or $\dot{a} $. Hence, all of these quantities would be of very much smaller order in the scenario of our interest, where the radial distance from the PBH is finite. The dominant term in the numerator of the ratio given in equation \ref{7.15} would be $H^{2}R$. Therefore, the overall ratio would have the order of $ \sim \, - \frac{2}{A^{2}}{\frac{H^{2}R}{H}} = - \frac{2}{A^{2}}HR   $ or $ \frac{2}{A^{2}}\dot{a} \tilde{r}$. As we are interested in the case where the radial distance from the PBH is finite, $ A \sim 1 $ and in the early radiation dominated era,\footnote{the only case where $\dot{a}$ may be near order 1 is the time just after the end of inflation} $ \dot{a} << 1 $, the resultant order of the ratio in equation \ref{7.15} is very smaller than 1. This implies that, in the scenario of our interest the quantity $  \Big( \frac{-2}{A^{2}} \frac{\partial C }{\partial \overline{t}} + \frac{2 C}{A^{3}} \frac{\partial A }{\partial \overline{t}}   \Big) $ may be neglected with respect to the quantity $ \Big( 1 - \frac{C^{2}}{A^{2}} \Big) \frac{1}{\Delta} \frac{\partial \Delta }{\partial \overline{t}} $. Again, we have previously shown that the quantity $ \Big( 1 - \frac{C^{2}}{A^{2}} \Big) \frac{1}{\Delta} \frac{\partial \Delta }{\partial \overline{t}}   $  is negligible in comparison with the quantity $\Big( 1 - \frac{C^{2}}{A^{2}} \Big)^{2} \frac{4F}{R} $. It is quite clear that the quantity $   \Big( 1 -  \frac{2C^{2}}{A^{2}} \Big)^{-1}  \left\lbrace \Big( 1 + \frac{C^{2}}{A^{2}} \Big) \frac{1}{A^{2}}  \frac{\partial C }{\partial \overline{t}} - \frac{2C}{A^{3}} \frac{\partial A }{\partial \overline{t}}  \right\rbrace \sim \, \Big( \frac{-2}{A^{2}} \frac{\partial C }{\partial \overline{t}} + \frac{2 C}{A^{3}} \frac{\partial A }{\partial \overline{t}}   \Big) $. Hence, the ratio in the equation \ref{7.12} is of order $ << 1 $. 
Thus, this analysis implies that the additive term in the equation \ref{7.2} with 1 can be neglected making the ratio  $ \mathbb{F}/\mathscr{F} \approx 1 $ or, $  \mathbb{F} \approx \mathscr{F} $. Again, in the quantity $\frac{  \mathbb{F} + i \sigma }{\mathscr{F} + i \sigma} $, if the frequency of the mode is not too small (i.e. if we neglect the ultra-low frequency modes), then too, the quantity $\frac{  \mathbb{F} + i \sigma }{\mathscr{F} + i \sigma}  $ would be $\approx 1 $. 
\vspace{0.5cm}\\
\textbf{Acknowledgement : }
Arnab Sarkar is grateful to Dr. Sumanta Chakraborty, for very useful discussions and his valuable suggestions, which were pivotal for overall improvement of this work. Arnab Sarkar also wants to thank Prof. Sayan Kar for various discussions related to this work, leading to clarification of several issues.  
Arnab Sarkar thanks S. N. Bose National Centre for Basic Sciences, Kolkata-106, India, under Department of Science and Technology, Govt. of India, for funding through institute-fellowship.    

\end{small}
\end{document}